\documentclass[%
 reprint,
 amsmath,
 amssymb,
 aps,
 nofootinbib,
 prd
]{revtex4-2}

\usepackage{graphicx}
\usepackage{dcolumn}
\usepackage{bm}

\usepackage[T1]{fontenc}
\usepackage[utf8]{inputenc}
\usepackage{amsmath}
\usepackage{mathtools} 
\usepackage{cancel} 
\usepackage{MnSymbol} 
\usepackage{graphicx,adjustbox} 
\usepackage{orcidlink} 
\usepackage{mathrsfs} 
\usepackage{simpler-wick} 
\usepackage{comment}
\usepackage{autobreak}
\usepackage[noend]{algpseudocode}
\usepackage{relsize}
\usepackage{tikz}
\usepackage{float}
\usepackage{stackengine}
\usetikzlibrary{decorations.pathmorphing}
\allowdisplaybreaks
\numberwithin{equation}{section}
\usepackage{nicematrix}
\usepackage{hyperref}
\hypersetup{
	colorlinks=true,
	linktoc=page,
	citecolor=americanrose,
	linkcolor=cadmiumgreen,
	urlcolor=blue} 
\definecolor{gr}{gray}{0.7}
\newcommand{\gr}[1]{{\color{gr}#1}}
\newcommand{\mzero}{\gr{\cdot}}
\newcommand{\PM}{\text{PM}}

\usepackage{xcolor}

\definecolor{americanrose}{rgb}{1.0, 0.01, 0.24}
\definecolor{cadmiumgreen}{rgb}{0.0, 0.42, 0.24}

\makeatletter
\newlength{\apb@width}
\newcommand{\autoparbox}[2][c]{\settowidth{\apb@width}{#2}\parbox[#1]{\apb@width}{#2}}
\newcommand{\includegraphicsbox}[2][]{\autoparbox{\includegraphics[#1]{#2}}}

\makeatletter\g@addto@macro\bfseries{\boldmath}\makeatother

\newcommand{\mysumint}{%
  \stackinset{c}{0.5pt}{c}{0pt}{$\scalebox{1.4}[2]{$\int$}$}{$\scalebox{1.3}[1.4]{$\sum$}$}
}

\def\e{\mathrm{e}}
\def\D{\mathrm{D}}
\def\d{\mathrm{d}}

\newcommand{\eps}{\epsilon}

\newcommand{\oldornew}[1]{}

\begin{document}

\title{\boldmath Gravitational Compton scattering at the fourth post-Minkowskian order}

\author{Giacomo Brunello$^{a,b,\,\orcidlink{0009-0004-4788-738X}}$}
\email{giacomo.brunello@sns.it}
\author{Mario Meo$^{a \,\orcidlink{0009-0005-5946-0487}}$}
\email{mario.meo@sns.it}
\author{Sid Smith$^{c,d,e,\, \orcidlink{0009-0007-7799-0136}}$}
\email{sid.smith@ed.ac.uk}

\affiliation{$^a$Scuola Normale Superiore, Piazza dei Cavalieri 7, 56126, Pisa, Italy}
\affiliation{$^b$INFN Sezione di Pisa, Largo
Pontecorvo 3, 56127 Pisa, Italy}
\affiliation{$^c$Dipartimento di Fisica e Astronomia, Università di Padova, Via Marzolo 8, 35131 Padova, Italy}
\affiliation{$^d$INFN, Sezione di Padova,
Via Marzolo 8, I-35131 Padova, Italy.}
\affiliation{$^e$Higgs Centre for Theoretical Physics, University of Edinburgh, James Clerk Maxwell Building, Peter Guthrie Tait Road, Edinburgh, EH9 3FD, United Kingdom}

\date{\today}

\begin{abstract}
We compute the classical gravitational Compton amplitude at the fourth post-Minkowskian order, $\mathcal{O}(G^4)$, 
within the Worldline Quantum Field Theory framework. We derive the associated $N$-matrix element, which provides the gravitational-wave scattering phase shift at the same order. As a nontrivial check, we show that our result agrees with black-hole perturbation theory.
\end{abstract}

\maketitle
\section{Introduction}
\begin{figure}[t]
  \centering
 \includegraphicsbox[width=0.4\textwidth]{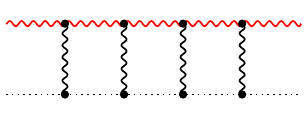}\>
  \caption{Three-loop integral topology appearing in the computation of the gravitational Compton amplitude at fourth PM order.
  Dotted lines represent the compact-object worldline, black wavy lines are potential
  gravitons, and the red wavy lines are the active graviton propagators.}
  \label{fig:three-loop-topology}
\end{figure}
Precise detection of gravitational-wave (GW) signals requires a deep understanding of the compact objects that generate them.
In particular,
it is important to characterise the response of compact objects
to external perturbations.
This response includes conservative tidal deformations,
dissipative absorption effects,
and more general frequency-dependent effects.
For neutron stars,
these quantities depend on the equation of state, whereas for black holes, they are fixed predictions of general relativity.

The traditional framework for computing black hole response is black hole perturbation theory (BHPT), which plays a central role in the description of wave propagation
in the presence of black holes.
For a Schwarzschild black hole,
linear gravitational perturbations are governed by the Regge–Wheeler
and Zerilli equations~\cite{Regge:1957td,Zerilli:1970se}.
Perturbations of rotating black holes are instead described
by the Teukolsky equation~\cite{Teukolsky:1973ha}.
The asymptotic solutions of these wave equations determine
partial-wave scattering phase shifts and absorption coefficients.
In the low-frequency regime,
these quantities can be computed systematically
using analytic and numerical methods,
including the Mano–Suzuki–Takasugi formalism~\cite{Mano:1996vt}
and related approaches~\cite{Dolan:2007mj}.
BHPT therefore provides the general-relativistic answer
to wave scattering off a black-hole background.

In parallel,
ideas and methods from quantum field theory have 
led to a new approach
to the relativistic two-body problem in general relativity,
combining effective field theory (EFT)~\cite{Goldberger:2004jt,Goldberger:2005cd},
scattering amplitudes,
and worldline methods.
These techniques have been predominantly applied across two separate perturbative schemes: The post-Minkowskian expansion for weak-field relativistic scattering between two compact objects and the post-Newtonian expansion for weak-field low-velocity bound systems.
In the post-Newtonian approach,
the combination of EFT ideas~\cite{Goldberger:2004jt,Goldberger:2005cd,Foffa:2013qca}
and multi-loop integration techniques~\cite{Foffa:2016rgu}
has pushed the study of conservative binary dynamics to 4PN~\cite{Foffa:2019rdf,Foffa:2019yfl}
and 5PN orders~\cite{Foffa:2019hrb,Porto:2024cwd,Porto:2026fsd},
and has recently led to partial results at 6PN~\cite{Brunello:2025gpf}
and 7PN~\cite{Brunello:2026anu}.
In the post-Minkowskian approach,
the computation of the 2PM gravitational Hamiltonian
from scattering amplitudes~\cite{Cheung:2018wkq,Bjerrum-Bohr:2018xdl}
initiated a new precision programme for conservative binary dynamics.
This programme has since reached 3PM~\cite{Bern:2019nnu,Bern:2019crd,Parra-Martinez:2020dzs,Cheung:2020gyp,DiVecchia:2021bdo,Brandhuber:2021eyq,Kalin:2020fhe},
4PM~\cite{Dlapa:2021npj,Bern:2021dqo,Bern:2022jvn,Dlapa:2022lmu,%
Jakobsen:2023ndj,Jakobsen:2023hig,Damgaard:2023ttc},
and more recently, the 5PM order~\cite{Bern:2023ccb,Driesse:2024xad,Bern:2024adl,%
Driesse:2024feo,Bern:2025wyd,Driesse:2026qiz,Dlapa:2026oyq}.
These results have been obtained using several complementary frameworks
for constructing the classical integrand,
including scattering-amplitude methods,
an observable-based approach~\cite{Kosower:2018adc},
worldline quantum field theory (WQFT)~\cite{Mogull:2020sak,Jakobsen:2021smu},
and worldline effective field theory~\cite{Kalin:2020mvi,Kalin:2022hph}.
A complementary line of research has focused on radiative observables,
using amplitudes and worldline methods to compute gravitational waveforms
at leading order~\cite{Jakobsen:2021smu,Mougiakakos:2021ckm,Jakobsen:2021lvp,DeAngelis:2023lvf,Brandhuber:2023hhl,Aoude:2023dui,Falkowski:2024bgb,Brunello:2025cot} 
and, more recently, at next-to-leading order~\cite{Brandhuber:2023hhy,Herderschee:2023fxh,%
Georgoudis:2023lgf,Elkhidir:2023dco,Caron-Huot:2023vxl,Bohnenblust:2023qmy,Brunello:2024ibk,Bohnenblust:2025gir,Brunello:2025eso}.
Most of these calculations treat compact objects as point-like sources and therefore do not include finite-size effects.

To incorporate such effects,
the compact object must be described 
by an effective worldline action
including non-minimal couplings to the gravitational field.
The Wilson coefficients of these operators encode the response
of the compact object to external perturbations.
Matching these coefficients to BHPT data is therefore a key step
toward a systematic description of compact-object response
within the EFT framework.

The object that naturally implements this matching
is the gravitational Compton amplitude,
also known as the gravitational Raman amplitude,
which describes the elastic scattering of a gravitational wave
off a compact object.
In the low-frequency regime,
this amplitude admits a post-Minkowskian expansion in powers of $Gm\omega$,
where $G$ is Newton’s constant,
$m$ is the mass of the compact object,
and $\omega$ is the frequency of the incoming radiation.
This quantity has been evaluated at 2PM order~\cite{Bjerrum-Bohr:2025bqg},
including spin effects~\cite{Akpinar:2025byi},
and recently at 3PM order for a non-spinning compact object~\cite{Bjerrum-Bohr:2026fhs,Bautista:2026qse,Ivanov:2026icp}.

In order to match these predictions with BHPT one needs to consider 
the exponential representation of the scattering matrix~\cite{Damgaard:2021ipf,DiVecchia:2023frv},
and extract the associated $N$-matrix,
also referred to as the Magnusian operator~\cite{Kim:2025gis}.
This relation has been made explicit for gravitational-wave scattering
in Refs.~\cite{Bjerrum-Bohr:2026fhs,Bautista:2026qse,Ivanov:2026icp},
where the exponentiated form of the amplitude was matched 
to the BHPT phase shift.

This perspective has recently played an important role
in the study of black-hole Love numbers,
dynamical response,
and finite-size effects in compact-object scattering~\cite{Ivanov:2022qqt,Saketh:2023bul,Ivanov:2024sds,Caron-Huot:2025tlq}.

Complementarily,
the Born-series approach to gravitational amplitudes provides
a direct partial-wave framework for matching effective theories to BHPT and for separating long-distance gravitational iterations
from short-distance tidal response~\cite{Correia:2024jgr,Caron-Huot:2025tlq}.

In this work,
we compute for the first time 
the minimal spinless gravitational Compton amplitude
at fourth post-Minkowskian order in WQFT.
The compact object is represented by a worldline coupled
to the gravitational field,
and classical correlators are generated
by Feynman diagrams with retarded boundary conditions ($+i\varepsilon$)~\cite{Jakobsen:2022psy,Kalin:2022hph}.

For wave scattering,
the central object is the graviton two-point function
in the presence of the spinless worldline source.
We construct the corresponding three-loop WQFT integrand,
reduce the resulting tensor structures to scalar integrals,
and map the amplitude to a three-loop integral family.

The required integrals are then evaluated using standard multi-loop techniques.
We use integration-by-parts identities
~\cite{Tkachov:1981wb,Chetyrkin:1981qh,Laporta:2000dsw}
to reduce the amplitude to a basis of 15 master integrals.
The master integrals are computed with the method of differential equations
~\cite{Kotikov:1990kg,Kotikov:1991pm,Remiddi:1997ny,Gehrmann:1999as,Argeri:2007up}
in the angular variable,
and the system is brought to canonical form
~\cite{Henn:2013pwa,Argeri:2014qva}.
The canonical system contains an elliptic sector, and the answer is naturally expressed in terms of elliptic iterated integrals.
We check the result by verifying the expected infrared-divergence structure.
Finally, we construct the fourth-order $N$-matrix operator by subtracting the lower-order Born iterations. We checked our result explicitly by matching it with BHPT.
\paragraph*{Conventions}

We work in mostly-minus metric, with signature $(+,-,-,-)$. Multi-loop integrals are defined in dimensional
regularization: $\D=4-2\epsilon$, with $\epsilon$ the analytic regulator.
For the loop integration measure we use the short-hand notation
\begin{align}
  \int_{\ell_1\ldots \ell_L}
  \equiv
\mu^{2\epsilon L} \frac{e^{\epsilon\, \gamma_E L}}{(4 \pi)^{\epsilon L}} \prod_{r=1}^{L}
  \int\frac{\d^\D\ell_r}{(2\pi)^\D}\,.
\end{align}
Here, $\mu$ is an arbitrary mass scale coming from dimen-
sional regularization, and $\gamma_E$ is the Euler–Mascheroni constant.

We denote spatial vectors in boldface, e.g. $\boldsymbol{\ell}$, 
and we use the notation
\begin{align}
  \ell^\mu = (\ell^0,\boldsymbol{\ell})\,,
\end{align}
where $\boldsymbol{\ell}$ is the $(\D-1)$-dimensional spatial part of the vector $\ell^\mu$. With this we also define the short-hand notation
\begin{align}
    \int_{\boldsymbol{\ell}_{1}\dots\boldsymbol{\ell}_{L}} \equiv \mu^{2\epsilon L}\frac{\e^{\epsilon\gamma_{E}L}}{(4\pi)^{\epsilon L}}\prod_{r=1}^{L}\int\frac{\d^{\D-1}\boldsymbol{\ell}_{r}}{(2\pi)^{\D-1}}\,.
\end{align}
We absorb the factor of $2\pi$ that appears in the worldline delta functions as
\begin{align}
  \hat\delta(y)
  \equiv
  2\pi\delta(y)\,.
\end{align}
\section{Setting up the problem}
\label{sec:setup}
\begin{figure}[!ht]
    \centering
    \includegraphicsbox[width=0.4\textwidth]{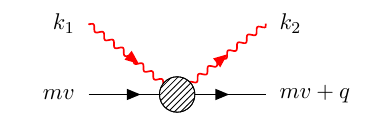}\>
    \caption{
      Kinematics of the problem. The red wavy lines describe the incoming $k_1$ and outgoing $k_2$ gravitational wave,
      while the black line is the worldline of the compact object of mass $m$ and four-velocity $v$. The momentum transfer $q=k_1-k_2$.
      \label{fig:compton_kin}
    }
\end{figure}
We consider the elastic scattering of a gravitational wave
off a spinless compact object of mass $m$,
as depicted in Fig.~\ref{fig:compton_kin}.
This process is described by the graviton two-point function
in the presence of a single massive worldline source,
and is commonly referred to as the gravitational Compton,
or gravitational Raman, amplitude.

The compact object is characterised by its gravitational charge
$r_s= Gm$, while the incoming radiation has wavelength
$\lambda_{\rm rad}= \omega^{-1}$, where $\omega$ is the frequency of the incoming and outgoing radiation. 
We work in the low-frequency regime
\begin{align}
  Gm\omega \ll 1 \, ,
  \qquad
  r_s \ll \lambda_{\rm rad}\, .
\end{align}
Under these assumptions, the gravitational wave probes only the long-distance field of the compact object. The latter can therefore be described by an effective worldline
coupled to gravity.
Equivalently, physical observables can be computed in a
post-Minkowskian expansion in powers of
\begin{equation}
\label{eq:PM_def}
    \epsilon_{\PM} = 2Gm\omega \,.
\end{equation}
\subsection{Kinematics}
The incoming graviton has momentum $k_1^\mu$
and helicity $\lambda_1$,
while the outgoing graviton has momentum $k_2^\mu$
and helicity $\lambda_2$.
We define the momentum transfer as
\begin{align}
  q^\mu
  =
  k_1^\mu-k_2^\mu \, .
\end{align}
The compact object has momentum $p^\mu = m v^\mu$, 
where $v^\mu$ is the classical four-velocity of the worldline,
normalised as $v^2=1$. 
The external gravitons are on-shell
\begin{align}
  k_1^2=k_2^2=0 \,.
\end{align}
Working in the rest frame of the compact object, we have $v=(1,\mathbf{0})$. As we will see, worldline Feynman rules produce an overall factor $\hat{\delta}(v\cdot q)$, enforcing conservation of the energy measured along the worldline. Therefore, the process is elastic
\begin{align}
  v\cdot k_1
  =
  v\cdot k_2
  \equiv
  \omega\, , \quad v\cdot q = 0\,.
\end{align}
We are also free to parametrize the graviton momenta in terms of the scattering angles in spherical coordinates $(\theta,\phi)$ as
\begin{align}
    k_{1} &= \omega(1,0,0,1)\, ,\nonumber\\
    k_{2} &= \omega(1,\sin\theta\cos\phi,\sin\theta\sin\phi,\cos\theta)
    \, ,\\
    q^{2} &= -4\omega^2 \sin^2\frac{\theta}{2}\equiv -4\omega^{2}x^{2}\,,\nonumber
\end{align}
where we have introduced the dimensionless variable
\begin{equation}
x=\sin\frac{\theta}{2}\, , \qquad 0\leq  x \leq 1 \, .
\end{equation}

Finally, the polarisation tensors are transverse and traceless,
\begin{align}%
  k_{i\mu}\varepsilon^{\mu\nu}_{i,\lambda_i}=0\, ,
  \qquad
  \eta_{\mu\nu}\varepsilon^{\mu\nu}_{i,\lambda_i}=0\, .
\end{align}
We opt to work in the transverse traceless (TT) gauge, in which the external polarizations are transverse to the worldline velocity
\begin{align}
\label{eq:gauge}
  v_\mu\varepsilon^{\mu\nu}_{i,\lambda_i}=0\, .
\end{align}
The graviton polarisation tensors can also be rewritten in a double-copy form, as products of spin-1 polarisation vectors as 
\begin{align}
    \varepsilon^{\mu\nu}_{i,\lambda_i}=\varepsilon^\mu_{i,\lambda_i}\varepsilon^\nu_{i,\lambda_i}\,.
\end{align}
We can parametrise the polarisation vectors in the rest frame as
\begin{align}
    \varepsilon_{1,\lambda_{1}} =&\, \frac{1}{\sqrt{2}}(0,1,0,0)+\frac{i\lambda_{1}}{\sqrt{2}}(0,0,1,0)\,,\nonumber\\
    \varepsilon_{2,\lambda_{2}} =&\, \frac{1}{\sqrt{2}}(0,\cos\theta\cos\phi,\cos\theta\sin\phi,-\sin\theta)\label{eq:polarizations}\\
    &+\frac{i\lambda_{2}}{\sqrt{2}}(0,-\sin\phi,\cos\phi,0)\,.\nonumber
\end{align}
\subsection{Worldline QFT Setup}
\label{sec:wqft-action}
The computation is formulated in the worldline quantum field theory (WQFT) framework~\cite{Mogull:2020sak}. 
The system is described by the action
\begin{align}
  S
  =
  S_{\rm EH}
  +
  S_{\rm GF}
  +
  S_{\rm pp} \, .
\end{align}
The bulk dynamics are governed by the Einstein-Hilbert action
\begin{align}
  S_{\rm EH}
  =
  -\frac{2}{\kappa^2}
  \int d^D x\,\sqrt{-g}\,R \, ,
\end{align}
and we use the de Donder gauge-fixing term
\begin{align}
  S_{\rm GF}
  =
  \frac{1}{\kappa^2}
  \int d^D x\,\sqrt{-g}\,g^{\mu\nu}\Gamma_\mu \Gamma_\nu \, ,
\end{align}
where $\Gamma_\mu = g^{\rho\sigma}\Gamma_{\rho\sigma\mu}$ and $\Gamma_{\rho\sigma\mu}$ are the Christoffel symbols, and $\kappa^{2}=32\pi G$ is the gravitational coupling constant.

The compact object is described by the gauge-fixed Polyakov form of the massive point-particle action~\cite{Mogull:2020sak}
\begin{align}
  S_{\rm pp}
  =
  -\frac{m}{2}
  \int d\tau\,
  g_{\mu\nu}(x(\tau))
  \dot{x}^{\mu}(\tau)
  \dot{x}^{\nu}(\tau) \, .
\end{align}
Here $x^\mu(\tau)$ is the worldline position and
$\tau$ is the worldline time parameter.
Dots denote differentiation with respect to $\tau$.
We can study this system perturbatively by expanding the metric around flat spacetime
\begin{align}
  g_{\mu\nu}
  =
  \eta_{\mu\nu}
  +
  \kappa \, h_{\mu\nu} \, ,
\end{align}
and the worldline around the straight trajectory
of the compact object
\begin{align}
  x^\mu(\tau)
  =
  v^\mu \tau
  +
  z^\mu(\tau) \, .
\end{align}
Here, $v^\mu$ is the constant asymptotic four-velocity and $z^\mu(\tau)$ denotes the worldline fluctuation.
Substituting these expansions into the action generates
the WQFT Feynman rules.
The point-particle action produces worldline vertices
coupling a single graviton to the compact object,
as well as vertices involving the fluctuation field $z^\mu$.
The Einstein--Hilbert action produces the usual bulk graviton
self-interactions.
In momentum space,
each integration over the straight worldline gives a delta function
enforcing the conservation of energy measured along the worldline.

We adopt the causal, or in-in, prescription of WQFT, based on the Schwinger–Keldysh formalism for real-time dynamics~\cite{Schwinger:1960qe,Keldysh:1964ud}. The diagrammatic rules therefore employ \textit{active propagators} with retarded $i\,\varepsilon$ prescription, which enforce causal propagation from the worldline source to the radiation field~\cite{Jakobsen:2022psy,Kalin:2022hph}
\begin{align}
    \vcenter{\hbox{\includegraphics[width=0.1\textwidth]{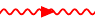}}}
 &\  =\  \frac{i \mathcal{P}_{\mu\nu\rho\sigma}}{(k_0+i\varepsilon)^2-\mathbf{k}^2}\,,\nonumber
    \\ \vcenter{\hbox{\includegraphics[width=0.1\textwidth]{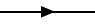}}}  &\ = -\ \frac{i \eta_{\mu\nu}}{m(\omega+i\varepsilon)^2}\,,
\end{align}
where $\mathcal{P}_{\mu\nu\rho\sigma}$ reads as
\begin{align}
    \mathcal{P}_{\mu\nu\rho\sigma} \ = \ \frac{1}{2}\left(\eta_{\mu\rho}\eta_{\nu\sigma}
    +\eta_{\mu\sigma}\eta_{\nu\rho}-\frac{2}{D-2}\eta_{\mu\nu}\eta_{\rho\sigma}\right)\,.
\end{align}
We define the transfer operator $\hat{T}$ as the connected part of the $\hat{S}$-matrix
\begin{align}
    \hat{S}= \mathbf{1}+i\, \hat{T} \,.
\end{align}
The gravitational Compton (or Raman) amplitude is then the matrix element of $i\hat{T}$ between asymptotic one-graviton states $\vert k_i,\lambda_i\rangle$
\begin{align}
  \langle k_2,\lambda_2 | i\hat T | k_1,\lambda_1\rangle
  =
  \hat{\delta}(v\cdot q)\,
  i\mathcal{M}(\lambda_1,\lambda_2) \, .
\end{align}
The amplitude admits a perturbative expansion in the post-Minkowskian parameter
\begin{align}
\label{eq:amplitude_expansion}
   i  \mathcal{M}(\lambda_1,\lambda_2) &= \sum_{n\ge1}   \mathcal{M}^{(n)}(\lambda_1,\lambda_2)\,,\nonumber\\
   \mathcal{M}^{(n)}(\lambda_{1},\lambda_{2})&= \mathcal{O}(\epsilon_\PM^n)\,.
\end{align}
In this work, we evaluate the contribution to the fourth post-Minkowskian order, $\mathcal{M}^{(4)}(\lambda_{1},\lambda_{2})$.
\section{Calculation of the Compton Amplitude}
\label{sec:recursive-integrand-generation}

In this section we will describe the calculation of the fourth post-Minkowskian contribution to the gravitational Compton amplitude.

\subsection{Recursive integrand generation}
To construct the WQFT integrand for the gravitational Compton amplitude introduced in Sec.~\ref{sec:wqft-action}, we consider the connected graviton two-point function in the presence of the spinless worldline source
\begin{align}
  \mathcal{G}_{\mu\nu\rho\sigma}(x_2,x_1)
  =
  \big\langle h_{\mu\nu}(x_2)h_{\rho\sigma}(x_1)\big\rangle_{\rm conn.} \,.
  \label{eq:2pt-def}
\end{align}
In the WQFT path-integral representation this reads as
\begin{align}
  \mathcal{G}_{\mu\nu\rho\sigma}
  =
  \frac{1}{\mathcal{Z}_{}}
  \int \mathcal{D}[h,z]
  h_{\mu\nu}(x_2)h_{\rho\sigma}(x_1)\,
  e^{iS[h,z]}\,.
  \label{eq:2pt-path-integral}
\end{align}
After Fourier transforming to momentum space and amputating the two external graviton propagators, we obtain an amputated momentum space kernel, $\widetilde{\mathcal{G}}^{\,\rm amp}_{\mu\nu\rho\sigma}(k_2,k_1;v)$, which is related to the amplitude as 
\begin{align}
  \hat{\delta}(v\cdot q)\,
  i\mathcal{M}(\lambda_1,\lambda_2)
  =
  \left(\varepsilon^{\mu\nu}_{2,\lambda_2}\right)^{*}
  \widetilde{\mathcal{G}}^{\,\rm amp}_{\mu\nu\rho\sigma}\,
  \varepsilon^{\rho\sigma}_{1,\lambda_1}\,,
  \label{eq:G-to-M}
\end{align}
The compact object is not a fixed background geometry; its long-distance field is built perturbatively by worldline insertions and bulk graviton self-interactions. In the absence of the incoming wave, the spinless worldline sources a static one-point function
\begin{align}
  H_{\mu\nu}(x)
  \equiv
  \big\langle h_{\mu\nu}(x)\big\rangle_{\rm WQFT}\,.
  \label{eq:H-def}
\end{align}
This is the WQFT representation of the Schwarzschild field in the chosen gauge, and it admits a $Gm$--expansion
\begin{align}
  H_{\mu\nu}
  =
  \sum_{r\geq1} H_{\mu\nu}^{(r)}\,,
  \qquad
  H_{\mu\nu}^{(r)}
  =
  \mathcal{O}\big((Gm)^r\big)\, .
  \label{eq:H-expansion}
\end{align}
Diagrammatically, $H_{\mu\nu}$ is a one-point current: at leading order a single graviton is emitted from the straight worldline, and higher orders attach further worldline sources through bulk self-interactions. It is thus the long-distance field generated by the compact object itself,
\begin{align}
\vcenter{\hbox{\includegraphics[height=1.25cm]{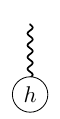}}}  = \vcenter{\hbox{\includegraphics[height=1.25cm]{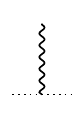}}}  +\,  \frac{1}{2}  \vcenter{\hbox{\includegraphics[height=1.25cm]{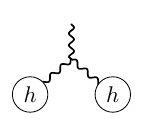}}}  +\,  \frac{1}{6}    \vcenter{\hbox{\includegraphics[height=1.2cm]{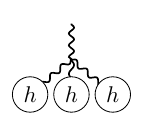}}}  +\,  ... \, .
\end{align}
The wave-scattering problem is the linear response of this configuration: $\mathcal{G}_{\mu\nu\rho\sigma}$ describes an active graviton entering at $x_1$, propagating through the field $H_{\mu\nu}$, and leaving at $x_2$. It is generated by a Berends–Giele~\cite{Berends:1987me} recursion (see Refs.~\cite{Jakobsen:2023ndj,Jakobsen:2023oow,Bautista:2026qse} for WQFT implementations), which we write schematically as
\begin{align}
  \mathcal{G}
  =
  \mathcal{G}_0
  +
  \mathcal{G}_0\,\mathcal{K}\,\mathcal{G}\,,
  \label{eq:dyson}
\end{align}
where $\mathcal{G}_0$ is the active flat-space graviton propagator and $\mathcal{K}$ is the interaction kernel seen by the active graviton. 
We split the kernel into a background and a recoil part
\begin{align}
  \mathcal{K}^{(r)}
  =
  \mathcal{K}_{\rm bg}^{(r)}
  +
  \delta_{r1}\,\mathcal{K}_{\rm rec}\,.
  \label{eq:K-split}
\end{align}
The background kernel $\mathcal{K}_{\rm bg}^{(r)}$ is the direct scattering of the active graviton off the order-$(Gm)^r$ field of the compact object, built from Einstein–Hilbert vertices with two active legs and any number of additional legs sewn into the one-point current $H_{\mu\nu}$
\begin{align}
\centering
\vcenter{\hbox{\includegraphics[height=2.5cm]{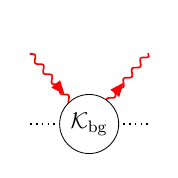}}}  = \sum_{n=1}^{\infty} \, \frac{1}{n!}\vcenter{\hbox{\includegraphics[height=2.2cm]{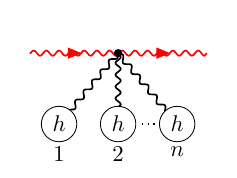}}}&\,. 
\end{align}
The recoil kernel is the single insertion carrying the worldline-fluctuation propagator: the active graviton excites a fluctuation $z^\mu$ through the worldline vertices, which propagates along the worldline with an active propagator, and is reabsorbed
\begin{align}
\centering
\vcenter{\hbox{\includegraphics[height=2.5cm]{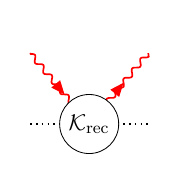}}}  = \vcenter{\hbox{\includegraphics[height=1.5cm]{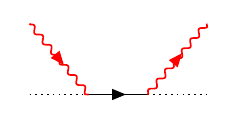}}}&\,.
\end{align}
The Kronecker delta in Eq.~\eqref{eq:K-split} reflects that recoil enters only as this leading insertion; higher-order recoil diagrams arise by iterating it together with the background kernels inside the recursion, not from new kernels.

Expanding the kernel and the connected two-point function perturbatively as
\begin{align}
  \mathcal{K}
  =
  \sum_{r\geq1}\mathcal{K}^{(r)}\,,
  \qquad
  \mathcal{G}
  =
  \sum_{n\geq0}\mathcal{G}^{(n)}\,,
  \qquad
  \mathcal{G}^{(0)}=\mathcal{G}_0 \,,
  \label{eq:G-K-expansion}
\end{align}
the recursion~\eqref{eq:dyson} with the kernel split~\eqref{eq:K-split} becomes
\begin{align}
\centering
\vcenter{\hbox{\includegraphics[height=1.2cm]{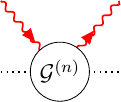}}}  = &\sum_{r=1}^{n}\vcenter{\hbox{\includegraphics[height=1.2cm]{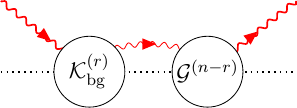}}}\nonumber \\
& 
\ \ +  \vcenter{\hbox{\includegraphics[height=1.2cm]{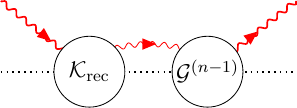}}}&\,,\label{eq:recursion}
\end{align}
which generates all connected WQFT diagrams with one incoming and one outgoing active graviton.

The diagrams generated by Eq.~\eqref{eq:recursion} have a simple causal structure. Potential gravitons, which carry vanishing energy in the rest frame of the compact object, build the static field $H_{\mu\nu}$, while active gravitons carry the external frequency $\omega$ and propagate the gravitational wave through this field. With retarded boundary conditions, causality flows from the incoming to the outgoing active graviton, and only active lines can go on shell.

Each worldline insertion produces a delta function $\widehat{\delta}(v\cdot Q)$, with $Q^\mu$ the total momentum entering the worldline segment, while bulk vertices impose ordinary $D$-dimensional momentum conservation. These constraints fix the energy flow along the worldline, so that the $n$PM contribution is an $(n-1)$-loop spatial integral.

The recursion generates $70$ WQFT diagrams at 4PM order, of which only $36$ contribute after imposing the external-polarisation gauge of Eq.~\eqref{eq:gauge}.
The diagram counts through 4PM order are collected in Tab.~\ref{tab:diagrams}.
\begin{table}[t]
    \centering
    \begin{tabular}{ccc}
        \hline\hline
        Loop order $L$ & PM order & Diagrams \\
        \hline
        0 & 1PM & $2\,(1)$ \\
        1 & 2PM & $6\,(3)$ \\
        2 & 3PM & $20\,(10)$ \\
        3 & 4PM & $70\,(36)$ \\
        \hline\hline
    \end{tabular}
    \caption{WQFT diagrams generated by the recursion~\eqref{eq:recursion}
    at loop order $L$, corresponding to PM order $n=L+1$. The numbers in parentheses denote the diagrams that survive after imposing
    $v_\mu\varepsilon^{\mu\nu}_{i,\lambda_i}=0$.}
    \label{tab:diagrams}
\end{table}
The Wick contractions, index contractions, and tensor simplifications needed to construct the momentum-space integrands were carried out with an in-house \texttt{FORM}~\cite{Davies:2026cci} implementation, using the on-shell conditions, the transversality and tracelessness of the external gravitons, the elastic relation $v\cdot q=0$, and the kinematic identities of Sec.~\ref{sec:setup}. As checks, we verified the external Ward identities and reproduced the lower-order WQFT results of Refs.~\cite{Bautista:2026qse, Bjerrum-Bohr:2026fhs,Ivanov:2026icp}.
\subsection{Integral reduction}
\label{sec:integral-reduction}

\begin{figure}[!t]
\includegraphics[width=0.2\textwidth]{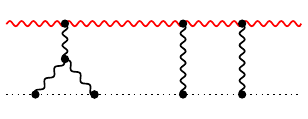}
\includegraphics[width=0.2\textwidth]{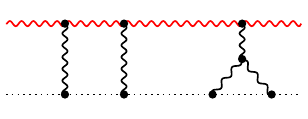}\\
\includegraphics[width=0.2\textwidth]{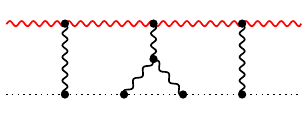}
\includegraphics[width=0.2\textwidth]{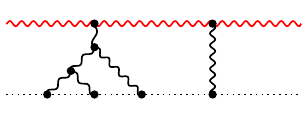}\\
\includegraphics[width=0.2\textwidth]{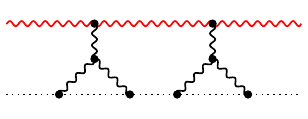}
\includegraphics[width=0.2\textwidth]{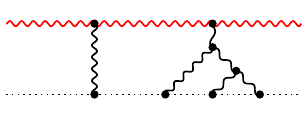}
   \caption{Reducible top sectors appearing in the three-loop Compton amplitude.}
  \label{fig:reducible-top-sectors}
\end{figure}
The recursive construction gives a sum of diagrammatic integrands with different momentum routings and worldline delta functions. The integrand is characterised by tensor integrals up to rank 4, where tensors contain polarisation vectors contracted with loop momenta $\varepsilon_i\cdot\ell_j$.

The tensors are reduced to scalar integrals using an in-house implementation of the Passarino-Veltman reduction in \texttt{FORM}, as described in Ref.~\cite{Anastasiou:2023koq}. All tensor dependence can be expressed in terms of three independent gauge-invariant structures~\cite{Bjerrum-Bohr:2025bqg}
\begin{align}
  \mathcal{T}_1
  &\equiv
  \bigl(v\cdot F_1\cdot F_2^*\cdot v\bigr)^2\,,\nonumber
  \\
  \mathcal{T}_2
  &\equiv
  \bigl(F_1\cdot F_2^*\bigr)
  \bigl(v\cdot F_1\cdot F_2^*\cdot v)\,,
  \\
  \mathcal{T}_3
  &\equiv
  (F_1\cdot F_2^*\bigr)^2\,,\nonumber
\end{align}
where $F_i^{\mu\nu}=k_i^\mu\epsilon_i^\nu-k_i^\nu\epsilon_i^\mu$ is the linearised field strength of the external graviton.
In this basis, the integrand can be written as a linear combination of the three tensor structures $\mathcal{T}_i$.
We map the resulting scalar integrals for all diagrams to a single class of integrals by employing an in-house topology-mapping algorithm in \texttt{FORM} that performs shifts of loop momenta to identify graphs by exploiting the structure of their associated Symanzik polynomials. The resulting integral family is given by
\begin{align}
    F_{n_{1}\dots n_{12}} = \int_{\ell_{1}\ell_{2}\ell_{3}}\frac{\prod_{r=1}^{3}\hat{\delta}(v\cdot\ell_{r})}{\prod_{a=1}^{12}D_{a}^{n_{a}}}\,,
\end{align}
where the propagators are
\begin{align}
    D_{a} &= (\ell_{a}^{0}+k_{1}^{0}+i\varepsilon)^{2}-(\mathbf{\ell}_{a}+\mathbf{k}_{1})^{2}\,,
   \nonumber \\
    D_{3+a} &= \ell_{a}^{2}\,,
   \nonumber \\
    D_{6+a} &= (\ell_{a}+q)^{2}\,,\label{eq:propagators_full}\\
    D_{9+a} &= (\ell_{a+1}-\ell_{a})^{2}\, , \quad \ell_{4}=\ell_{1}\,,\nonumber
\end{align}
where $a=1,2,3$. These integrals depend on two kinematic variables: the graviton frequency $\omega=v\cdot k_{i}$ and the angular variable $x$. There are 7 highest-level sectors appearing with $7$ denominators, of which only one is irreducible, shown in Fig.~\ref{fig:three-loop-topology}; the reducible ones are shown in Fig.~\ref{fig:reducible-top-sectors}.

We perform integration-by-parts (IBP) reduction using the code \texttt{PRISM}~\cite{PRISM}, based on syzygy-based methods~\cite{Gluza:2010ws,Wu:2023upw,Wu:2025aeg,Smith:2025xes}, improved seeding
algorithms~\cite{Lange:2025fba,Wu:2023upw,Wu:2025aeg}, and
finite-field reconstruction techniques as implemented in \texttt{FiniteFlow}~\cite{Peraro:2019svx}.

\begin{figure}[!ht]
\includegraphics[width=0.23\textwidth]{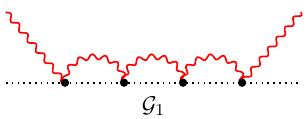}
\includegraphics[width=0.23\textwidth]{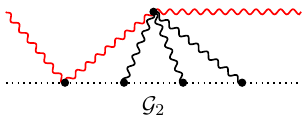}
\vspace{1mm}
\includegraphics[width=0.23\textwidth]{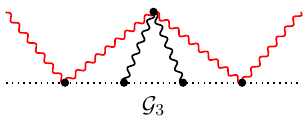}
\includegraphics[width=0.23\textwidth]{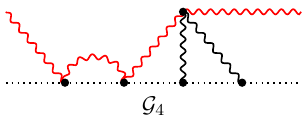}
\vspace{1mm}
\includegraphics[width=0.23\textwidth]{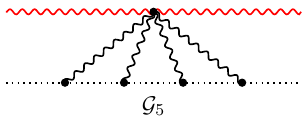}
\includegraphics[width=0.23\textwidth]{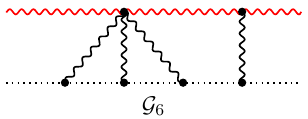}
\vspace{1mm}
\includegraphics[width=0.23\textwidth]{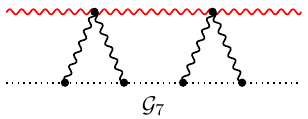}
\includegraphics[width=0.23\textwidth]{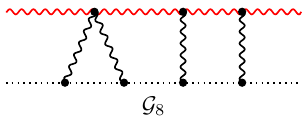}
\vspace{1mm}
\includegraphics[width=0.23\textwidth]{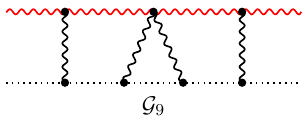}
\includegraphics[width=0.23\textwidth]{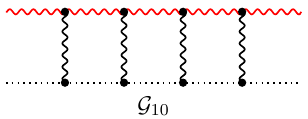}
   \caption{10 master integral topologies $\{\mathcal{G}_i\}_{i=1}^{10}$ 
   containing the 15 master integrals appearing in the three-loop integral reduction.
   The red wavy lines represent the active graviton propagators (see Sec.~\ref{ssec:integral_evaluation}), 
   while black wavy lines represent the potential graviton propagators. 
   Black dotted lines denote worldline delta functions.}
  \label{fig:three-loop-subgraphs}
\end{figure}

After IBP reduction, the amplitude can be expressed in terms of a basis of $15$ master integrals, spanning over $10$ sub-topologies, depicted in Fig.~\ref{fig:three-loop-subgraphs}, which are all sub-sectors of the irreducible top sector in Fig.~\ref{fig:three-loop-topology}. The number of master integrals for each sub-topology is given by the following set of integrals: 
\begin{align}
    \mathcal{G}_{1,\dots,6}: 1\, , \quad 
    \mathcal{G}_{7,8,9}:2\, , \quad 
    \mathcal{G}_{10}:3 \, .
\end{align}
Interestingly, we find that $7$ of the $36$ diagrams are zero after IBP reduction. These $7$ diagrams are the only ones containing a worldline propagator. We observed the same pattern at lower loops and therefore conjecture that this holds at all orders in the TT gauge.
\subsection{Integral evaluation}
\label{ssec:integral_evaluation}
In order to evaluate the master integrals, we solve the associated differential equations that the master integrals obey and fix the boundary conditions in the forward-scattering limit. 

\begin{figure}[!ht]
  \centering
\includegraphics[width=0.3\textwidth]{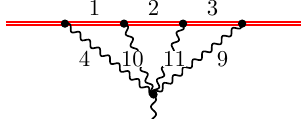}
  \caption{
  Three-loop master integral topology after imposing the worldline delta functions. 
  The red double lines represent massive propagators with mass $\omega$, 
  while black wavy lines represent massless propagators.}
  \label{fig:spatial-three-loop-family}
\end{figure}

We first note that the $i\, \varepsilon$ prescription dictated by the Feynman rules only has an effect on active propagators $D_{1,2,3}$, due to the worldline delta functions. This can be seen by imposing the delta functions, after which the family takes the form
\begin{align}
    F_{n_{1}\dots n_{12}}^{(\eta_{1}\eta_{2}\eta_{3})} = \int_{\boldsymbol{\ell}_{1}\boldsymbol{\ell}_{2}\boldsymbol{\ell}_{3}}\frac{1}{\prod_{a=1}^{3}\tilde{D}_{a}(\eta_{a})^{n_{a}}\prod_{a=4}^{12}\tilde{D}_{a}^{n_{a}}} \, ,
\end{align}
where
\begin{align}
    \tilde{D}_{a} &= (\omega+i\eta_{a}\varepsilon)^2-(\boldsymbol{\ell}_{a}+\boldsymbol{k}_{1})^{2}\,,
    \nonumber\\
    \tilde{D}_{3+a} &= -\boldsymbol{\ell}_{a}^{2}\,,
    \nonumber\\
    \tilde{D}_{6+a} &= -(\boldsymbol{\ell}_{a}+\boldsymbol{q})^{2}\,,
    \\
    \tilde{D}_{9+a} &= -(\boldsymbol{\ell}_{a+1}-\boldsymbol{\ell}_{a})^{2}\,, \quad \boldsymbol{\ell}_{4}=\boldsymbol{\ell}_{1}\,.\nonumber
\end{align}
In order to eventually calculate the phase shift associated to the scattering amplitude, we will need the integrals evaluated with \textit{advanced propagators} ($-i\varepsilon)$ as well, this is why we have included the prescriptions $\eta_{a}=\pm$.
The resulting irreducible top sector after imposing the delta function is depicted in Fig.~\ref{fig:spatial-three-loop-family}.
\subsubsection*{Differential Equations} 
We denote the set of master integrals by $\mathbf{J}$, and these satisfy a system of differential equations in the kinematic variable $x$\footnote{The dependence of the integrals on the graviton frequency $\omega$ is trivially found by dimensional analysis.}
\begin{align}
    \partial_{x}\mathbf{J}(x,\epsilon) = A(x,\epsilon)\mathbf{J}(x,\epsilon)\,.
\end{align}
A convenient choice of $\mathbf{J}$ is the canonical basis \cite{Henn:2013pwa}, where the dependence on the spacetime dimension parameter $\epsilon$ is factorised from the kinematic dependence
.
\begin{align}
    \partial_{x}\mathbf{J}(x,\epsilon) = \epsilon\hat{A}(x)\mathbf{J}(x,\epsilon)\,.
\end{align}
To find the canonical basis, we used a combination of methods involving leading singularities~\cite{Flieger:2022xyq,Gorges:2023zgv,Duhr:2025lbz,Forner:2026vby}, Magnus expansion~\cite{Argeri:2014qva}, and the publicly available code \textsc{Canonica}~\cite{Meyer:2017joq}. The explicit form of the canonical differential equation is reported in the ancillary file \texttt{de\_canonical.m}.

The system is characterised by a family of elliptic integrals that appears in the sector $\mathcal{G}_7$ of Fig.~\ref{fig:three-loop-subgraphs},
which is the solution of the following second-order differential operator:
\begin{align}
\left(\partial_x^2
-
\frac{3x^2-1}{x(1-x^2)}\,\partial_x
-
\frac{1}{1-x^2}\right) w_0 = 0 \, .
\end{align}
In the kinematic region $0\leq x\leq 1$, this is given by the first elliptic period
\begin{align}
    w_{0}(x) = \frac{2}{\pi}K(x^{2})\,.
\end{align}
Details on the canonical differential equation system are given in 
App.~\ref{app:canonical}.
The solution can be expressed in terms of iterated integrals~\cite{Chen:1977oja,Remiddi:1999ew}, 
which, in this case, are given by elliptic polylogarithms~\cite{Broedel:2018qkq,Duhr:2019tlz}
\begin{align}
    \mathbf{J}=&\biggl( \mathbf{1}+ \epsilon \int_0^x d x' \hat{A}(x')\nonumber\\
    &
    + \epsilon^2 \int_0^x d x' \hat{A}(x')\int_0^{x'} d x'' \hat{A}(x'') +\ldots\biggr)\mathbf{J}_b \,,
\end{align}
where $\mathbf{J}_b$ is a boundary vector at the kinematic point $x=0$.
The integration kernel consists of the logarithmic singularities $(x,x+1,x-1)$ and the elliptic periods $w_0, w_0'$.
\subsubsection*{Boundary conditions}
\begin{figure}[!ht]
    \centering
    \includegraphics[width=0.23\textwidth]{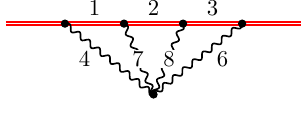}
    \includegraphics[width=0.23\textwidth]{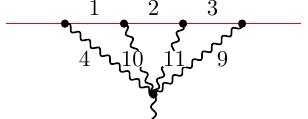}
    \caption{Boundary integral topologies in the hard (left) and soft (right) regions. In the hard region this is a three-loop two-point function with three massive propagators (red double lines). In the soft region we have a three-loop three-point function with three eikonal propagators (red single lines).}
    \label{fig:3loop_regions}
\end{figure}
For the boundary conditions, we analyse the behaviour of the master integrals in the forward-scattering limit
\begin{align}
	x\to0 \, .
\end{align}
In this limit, the differential equations develop a regular singular point, and the
asymptotic behaviour of the master integrals can be studied using the method of regions~\cite{Beneke:1997zp,Pak:2010pt,Jantzen:2012mw}.
At three loops, there are two relevant rescalings of the loop momenta as
\begin{align}
\text{hard: } \boldsymbol{\ell}_i \sim \omega \, ,
\qquad
\text{soft: } \boldsymbol{\ell}_i \sim \omega x  \, .
\nonumber
\end{align}
To minimise the number of integrals to compute, we study the asymptotic regions at the level of the 
differential equations~\cite{Mastrolia:2017pfy,Dulat:2014mda, Chestnov:2023kww,Brunello:2025cot}.
Such a strategy has already been studied in the two-loop case in Refs.~\cite{Bautista:2026qse,Ivanov:2026icp}.
The possible regions contributing are dictated by the eigenvalues of the residue matrix 
of the system at $x=0$
\begin{align}
  \epsilon\, \hat{A}(x,\epsilon)
  \overset{x\to 0}{=}
  \epsilon\frac{\hat{A}_0(\epsilon)}{x}
  +
  \mathcal{O}(x^0) \, .
\end{align}
In particular, at three loops, we can perform a Jordan decomposition
of the residue matrix, finding 3 regions
\begin{align}
	M\cdot \epsilon \hat{A}_0 \cdot M^{-1}
  =
  \text{Diag}
  \bigl[
  0^{\times 6},
  (-6\epsilon)^{\times 6},
  (2\epsilon)^{\times 3}
  \bigr]\,,
\end{align}
Here, the eigenvalue $0$ corresponds to the hard region, where all the rescalings are hard, while $(-6 \epsilon)$ corresponds to the soft region, where all the rescalings are soft.
The eigenvalue $(2 \epsilon)$ represents a spurious region, 
where all integrals become scaleless, so we impose that the integrals vanish in that region. Energy conservation imposes that there are no regions with mixed rescalings, which give rise to scaleless integrals.
Hence, the generic boundary vector is given by
\begin{align}
  \mathbf{J}_b = \mathbf{J}_0 + x^{-6\epsilon}\,\mathbf{J}_{-6\epsilon}\,.
\end{align}
where $\mathbf{J}_0$ and $\mathbf{J}_{-6\epsilon}$ are the boundary vectors for the hard and soft regions, respectively.
The new basis of master integrals $\mathbf{J}'=M\cdot \mathbf{J}$ diagonalizes the 
asymptotic behaviour of the system at $x=0$.
For the two regions, we perform an expansion under the integral sign, 
leading to two distinct families of integrals, depicted in Fig.~\ref{fig:3loop_regions}.
\paragraph{Hard region.}

In the hard region, the integral family becomes a three-loop two-point function with three massive propagators:
\begin{align}
	F_{n_{1}\dots n_{9}}^{\rm h \, (\eta_1 \eta_2 \eta_3)}
	=
	\int_{\boldsymbol{\ell}_1\boldsymbol{\ell}_2\boldsymbol{\ell}_3}
  \frac{1}{\prod_{a=1}^{9}(\tilde{D}^{\rm h}_{a})^{n_a}}\, ,
\end{align}
where
\begin{align}
  \tilde{D}^{\rm h}_a(\eta_{a}) &= (\boldsymbol{\ell}_a+\boldsymbol{k}_1)^2-(\omega+i\eta_a\varepsilon)^2\,,
  \nonumber\\
  \tilde{D}^{\rm h}_{3+a} &= \boldsymbol{\ell}_{a}^{2}\,, \\
  \tilde{D}^{\rm h}_{6+a} &= (\boldsymbol{\ell}_{a+1}-\boldsymbol{\ell}_{a})^{2}\,,\quad\boldsymbol{\ell}_{4}=\boldsymbol{\ell}_{1}\,.\nonumber
\end{align}
Using IBPs, we find 6 master integrals appearing
\begin{align}
    & F_{111000000}^{\rm h\, (\eta_1 \eta_2 \eta_3)}\, , &
    & F_{100001110}^{\rm h\, (\eta_1 \eta_2 \eta_3)}\, , &
    & F_{101000110}^{\rm h\, (\eta_1 \eta_2 \eta_3)}\, , &\nonumber\\
    & F_{110001010}^{\rm h\, (\eta_1 \eta_2 \eta_3)}\, , &
    & F_{010101110}^{\rm h\, (\eta_1 \eta_2 \eta_3)}\, , &
    & F_{111101110}^{\rm h\, (\eta_1 \eta_2 \eta_3)}\, .\label{eq:mis_hard}
\end{align}
Such MIs have already appeared in the context of $g-2$ computations (see, e.g., \cite{Laporta:2001rc,Lee:2010ik}).
\paragraph{Soft region.} 

In the soft region, the integral family is that of a soft three-loop triangle with eikonal propagators
\begin{align}
	F_{n_{1}\dots n_{12}}^{\rm s \, (\eta_1 \eta_2 \eta_3)}=\int_{\boldsymbol{\ell}_1\boldsymbol{\ell}_2\boldsymbol{\ell}_3}
  \frac{1}{\prod_{a=1}^{12}(\tilde{D}^{\rm s}_{a})^{n_a}}\,,
\end{align}
where
\begin{align}
    \tilde{D}_{a}^{\rm s}(\eta_{a}) &= \boldsymbol{\ell}_{a}\cdot\boldsymbol{k}_{1}-i\eta_{a}\varepsilon\,,\nonumber\\
    \tilde{D}^{\rm s}_{3+a} &= \boldsymbol{\ell}_{a}^{2}\,,\nonumber\\
    \tilde{D}^{\rm s}_{6+a} &= (\boldsymbol{\ell}_{a}+\boldsymbol{q})^{2}\,,\nonumber\\
    \tilde{D}^{\rm s}_{9+a} &= (\boldsymbol{\ell}_{a+1}-\boldsymbol{\ell}_{a})^{2}\,,\quad\boldsymbol{\ell}_{4}=\boldsymbol{\ell}_{1}\,.
\end{align}

Using IBPs, we find 6 master integrals appearing
\begin{align}
    & F_{000100001110}^{\rm s\, (\eta_1 \eta_2 \eta_3)}\, , &
    & F_{001100001110}^{\rm s\, (\eta_1 \eta_2 \eta_3)}\, , &
    & F_{010100001110}^{\rm s\, (\eta_1 \eta_2 \eta_3)}\, , &\nonumber\\
    & F_{011100001110}^{\rm s\, (\eta_1 \eta_2 \eta_3)}\, , &
    & F_{101100001110}^{\rm s\, (\eta_1 \eta_2 \eta_3)}\, , &
    & F_{111100001110}^{\rm s\, (\eta_1 \eta_2 \eta_3)}\, .\label{eq:mis_soft}
\end{align}
Integrals of this type typically appear in the context of HQET (see, e.g.,\cite{Chetyrkin:2003vi}). Their analytic expressions have already been computed in Refs.~\cite{Dlapa:2023hsl,Jinno:2022sbr}.

Boundary integrals do not depend on any kinematic scale; for this reason, it is very convenient to evaluate them using numerical methods. For this task, we used \textsc{AMFlow}~\cite{Liu:2022chg} to compute all boundary integrals numerically up to at least 200 digits of precision. 
The analytic expressions were then reconstructed with a suitable choice of transcendental functions using the PSLQ algorithm implemented within \textsc{PolyLogTools}~\cite{Duhr:2019tlz}. 

We checked against known results in both regions. We also computed many of the integrals via Mellin-Barnes and direct integration techniques. Further details and analytic results are reported in Appendix~\ref{app:boundary}. The ancillary file \texttt{retarded\_boundary.m} contains the boundary integrals for the canonical basis up to $\mathcal{O}(\epsilon^{0})$ in the fully retarded case ($\eta_{i}=+$).
\section{Result}
The amplitude at all orders takes the following form
\begin{align}
    i\mathcal{M}(\lambda_1,\lambda_2) = \sum_{n=1}^{\infty}\sum_{k=1-n}^{\infty}\epsilon^{k}i\mathcal{M}_{k}^{(n)}(\lambda_{1},\lambda_{2})\,,
\end{align}
and contains infrared divergences as predicted by Weinberg~\cite{Weinberg:1965nx,Weinberg:1995mt,Weinberg:1996kr}.

We note that tree level has no $\epsilon$-dependence, and so $\mathcal{M}_{k>0}^{(1)}=0$. Infrared divergences exponentiate, and we can define the finite part of the amplitude as
\begin{align}\label{eq:finiteM}
    i\mathcal{M}_{\text{fin}}(\lambda_{1},\lambda_{2}) = \e^{i\epsilon_{\PM}/\epsilon}i\mathcal{M}(\lambda_{1},\lambda_{2})\,.
\end{align}
where $\epsilon_{\PM}$ is defined as in Eq.~\eqref{eq:PM_def}.
This finite part relates to the scattering cross-section and can be used to match with BHPT following the procedure outlined in Ref.~\cite{Bjerrum-Bohr:2026fhs}. 
The divergent part at all orders has the following structure
\begin{align}
    \mathcal{M}_{-k}^{(n)} = (-1)^{k}\sum_{m=1}^{n-k}\frac{(i\epsilon_{\PM})^{n-m}\binom{n-m-1}{k-1}}{(n-m)!}\mathcal{M}_{n-k-m}^{(m)}\,,
\end{align}
for $k>0$, which we have verified for $n\leq 4$. 
At fourth post-Minkowskian order, we have the structure
\begin{align}
    i\mathcal{M}^{(4)} =& \frac{i\mathcal{M}_{-3}^{(4)}}{\epsilon^{3}}+\frac{i\mathcal{M}_{-2}^{(4)}}{\epsilon^{2}}+\frac{i\mathcal{M}_{-1}^{(4)}}{\epsilon}+i\mathcal{M}_{0}^{(4)}\,\nonumber\nonumber \\
    =& -\frac{\epsilon_{\PM}^{3}\mathcal{M}_{0}^{(1)}}{6\epsilon^{3}}-\frac{i\epsilon_{\PM}^{2}\mathcal{M}_{0}^{(2)}}{2\epsilon^{2}}
    \nonumber\\
    &+\frac{1}{\epsilon}\left(\frac{i\epsilon_{\PM}^{2}}{2}\mathcal{M}_{1}^{(2)}+\epsilon_{\PM}\mathcal{M}_{0}^{(3)}\right)+i\mathcal{M}_{0}^{(4)} \, .
    \label{eq:amplitude_4pm}
\end{align}
The contributions $i\mathcal{M}_{k}^{(4)}$ are provided in the ancillary file \texttt{iM\_4pm.m}.

\subsection{Extracting the phase shift}
\label{sec:phase-shift}
To connect the 4PM scattering amplitude of Eq.~\eqref{eq:amplitude_4pm} with BHPT, we need to consider the exponential representation of the scattering matrix~\cite{Damgaard:2021ipf,Brandhuber:2025igz,Kim:2025gis}
\begin{align}
  \hat S=\e^{i\hat N}\,.
\end{align}
As shown in Refs.~\cite{Bautista:2026qse,Ivanov:2026icp}, the matrix elements of $\hat N$
\begin{align}
  \langle k_2,\lambda_2|i\hat N^{(n)}|k_1,\lambda_1\rangle
  =
  \hat{\delta}(v\cdot q)\,
  i\mathcal{N}^{(n)}_{\lambda_1\lambda_2}(x,\phi)\,.
\end{align}
are the objects that map directly onto the partial-wave phase shift. 
They are obtained
from the transfer matrix by expanding
\begin{align}
  i\hat N
  =
  \log(1+i\hat T)\,.
\end{align}
Both operators admit a PM expansion
\begin{align}
i\hat T = \sum_{n\geq 1} i\hat T^{(n)} \, ,
\qquad
i\hat N = \sum_{n\geq 1} i\hat N^{(n)} \, .
\end{align}
Inserting these expansions into the operator formula, we get
\begin{align}
i\hat N^{(4)}
& = 
i\hat T^{(4)}\nonumber\\
&-\frac{1}{2}
\bigl(
i\hat T^{(1)}\!\cdot i\hat T^{(3)}
+i\hat T^{(2)}\!\cdot i\hat T^{(2)}
+i\hat T^{(3)}\!\cdot i\hat T^{(1)}
\bigr)\nonumber\\
&+\frac{1}{3}
\bigl(
i\hat T^{(1)}\!\cdot i\hat T^{(1)}\!\cdot i\hat T^{(2)}+i\hat T^{(1)}\!\cdot i\hat T^{(2)}\!\cdot i\hat T^{(1)}\nonumber\\
&+i\hat T^{(2)}\!\cdot i\hat T^{(1)}\!\cdot i\hat T^{(1)}
\bigr)\nonumber \\
&-\frac{1}{4}\,
i\hat T^{(1)}\!\cdot i\hat T^{(1)}\!\cdot i\hat T^{(1)}\!\cdot i\hat T^{(1)} \, .\label{eq:N4_raw_log}
\end{align}
Enforcing unitarity perturbatively as done in Ref.~\cite{Ivanov:2026icp}
\begin{align}
    T-T^\dagger = i T T^\dagger \, ,
\end{align}
we obtain at 4PM
\begin{align}
i\hat N^{(4)}
&=
\text{Im}\left[i\hat T^{(4)}\right]
+\frac{1}{4} i\hat T^{(1)}\!\cdot  i\hat T^{(1)} \cdot i\hat  T^{(1)}\! \cdot i\hat T^{(1)} \nonumber \\
&
-\frac{1}{6}
\bigl(
i\hat T^{(2)}\! \cdot i\hat T^{(1)}\!\cdot  i\hat T^{(1)}
+
i\hat T^{(1)} \! \cdot i\hat T^{(2)}\!\cdot  i\hat T^{(1)}
\nonumber \\
& +
i\hat T^{(1)} \! \cdot\! i\hat T^{(1)}\!\cdot  i\hat T^{(2)}
\bigr) \,,
\label{eq:N4-subtractions}
\end{align}
where with $\cdot$ we denote the insertion of a complete set of intermediate on-shell graviton states
\begin{align}
  ( X_1\cdot\ X_2)_{fi}
  =
  \mysumint_{\lambda,\ell}
  \hat\delta_+(\ell^2)\,
  \langle f|\hat X_1|\ell,\lambda\rangle
  \langle \ell,\lambda|\hat X_2|i\rangle\,,
\end{align}
and for products with more factors, we have analogous multiple phase-space integrals.
The subtraction terms in Eq.~\eqref{eq:N4-subtractions} correspond to unitarity cuts of scattering amplitudes. 
Considering the propagator labelling of Eq.~\eqref{eq:propagators_full}, we have for the matrix element $\mathcal{N}^{(4)}$
\begin{align}
    i\mathcal{N}^{(4)}& = \textrm{Im}\left[i \mathcal{M}^{(4)}\right]-\frac{1}{4}\mathrm{Cut}_{(123)}\left[i\mathcal{M}^{(4)}\right]
    \nonumber \\
    & -\frac{1}{6}\bigl(\mathrm{Cut}_{(12)}\left[i\mathcal{M}^{(4)}\right]+\mathrm{Cut}_{(13)}\left[i\mathcal{M}^{(4)}\right]\nonumber\\
    & +\mathrm{Cut}_{(23)}\left[i\mathcal{M}^{(4)}\right]\bigr) \, .
\end{align}
The various cuts of the 3-loop scattering amplitudes $\mathrm{Cut}_{(...)}\left[i\mathcal{M}^{(4)}\right]$ are obtained by simply applying the associated cut at the level of the master integrals and substituting the corresponding cut propagators with on-shell delta functions normalised as $(2\pi i)\delta(D)$.
Hence, the matrix element $i\mathcal{N}^{(4)}$ can be obtained by considering cut-subtracted master integrals
\begin{align}
    \tilde{J}_i &= \mathrm{Re}[J_i] - \frac{1}{4}\mathrm{Cut}_{(123)} [J_i]
    \nonumber \\
    &-\frac{1}{6}\left(\mathrm{Cut}_{(12)}+\mathrm{Cut}_{(13)}+\mathrm{Cut}_{(23)}\right)[J_i] \, .
\end{align}
Furthermore, since the analytic dependence of the integrals on $x$ is invariant under cuts, it is sufficient to consider only cut-subtracted boundary conditions.
These have been computed numerically using \textsc{AMFlow}, and via reverse-unitarity~\cite{Anastasiou:2002yz,Anastasiou:2003gr}, rewriting delta functions as a linear combination of propagators with opposite $\pm i\, \varepsilon$ prescription
\begin{equation}
    (2 \pi i) \, \delta(p^2) = \frac{1}{(p^2-i\varepsilon)}-\frac{1}{(p^2+i\varepsilon)} \, .
\end{equation}
The explicit result for the cut-subtracted boundary vectors in the canonical basis that directly give the phase shift is available in the ancillary file \texttt{iterated\_boundary.m}.
The 4PM matrix elements for the helicity-conserving ($\lambda_{1}=\lambda_{2})$ and helicity-reversing $(\lambda_{1}=-\lambda_{2})$ cases are surprisingly simple and read as
\begin{widetext}
\begin{align}
    \mathcal{N}^{(4)}_{+ + }(x,\phi) & = \frac{\eps_{PM}^4\pi^2\e^{2i\phi}}{\omega(1-x^2)}\biggl[4 (7+11 x^2)\log(1+x)
   + \frac{2 x p_1(x) w_0'(x)+p_2(x) w_0(x)}{1536} \left(\frac{\pi^2}{4} -  \mathcal {G} _ 1 (x)+\frac{w_1(x)}{w_0(x)} \mathcal {G} _ 0 (x)\right)
   \nonumber\\ 
    &
   \quad  -\frac{1}{1-x^2}\biggl(\frac{4}{3} \pi ^2 p_3(x) +\frac{(1-x) p_4(x)}{23040}+16 p_3(x) \text{Li}_2(-x)
    +\frac{p_1(x)}{768} \frac {\mathcal {G} _ 0 (x)} {w_ 0 (x)}\biggr)\biggr] \, \label{eq:phase_4pm} ,
   \\ 
    \mathcal{N}^{(4)}_{+ - }(x,\phi) & = 0 \nonumber\, .
\end{align}
\end{widetext}
where
\begin{align}
    \mathcal{G}_i(x)&  = \int_0^x dx' w_i(x') \, , \qquad i=0,1\, ,
    \nonumber \\ 
    p_1(x)& 
    =17805 x^4+113738 x^2+44273 \, ,
    \nonumber\\
    p_2(x) & = 25485 x^4+120355 x^2+29976 \, ,
    \\
    p_3(x) & = 2 x^4+6 x^2+1 \, ,
    \nonumber\\
    p_4(x) & = 
    1575 x^6+47655 x^5+827547 x^4+366747 x^3 \, ,
    \nonumber \\
    &+3142262 x^2+166262 x+722432 \,\nonumber ,
\end{align}
and $w_i(x)$ are the elliptic periods
\begin{align}
    w_0(x) = \frac{2}{\pi}K(x^2) \, , \qquad w_1(x) = K(1-x^2) \, .
\end{align}
We notice that the result is finite and is expressed in terms of a small set of functions:
\begin{align}
   \left\{ \log(1+x),\  \text{Li}_2(-x), \ w_i(x), \ \mathcal{G}_i(x)\right\} \, . 
\end{align}
Eq.~\eqref{eq:phase_4pm} is also reported in the ancillary file \texttt{N\_matrix\_4pm.m}. We remark that no explicit computation of cuts of scattering amplitudes was needed.
\subsection{Matching with BHPT}
To match the matrix element of Eq.~\eqref{eq:phase_4pm} with BHPT, we followed the approach outlined in Ref.~\cite{Bautista:2026qse}.  The BHPT computation is naturally formulated in a basis of
spin-weighted spherical harmonics. For a Schwarzschild background, the two parity
sectors are described by the Regge--Wheeler and Zerilli radial equations. 
The large-radius solution defines
the partial-wave scattering matrix
\begin{align}
  {}_{-2}S^P_{\ell 2}
  =
  {}_{-2}\eta_{\ell 2}\,
  e^{2i\,{}_{-2}\delta^P_{\ell 2}}\,,
  \qquad
  P=\pm1\,,
\end{align}
where $P$ labels the even and odd parity channels. In the low-frequency expansion, the absorption factor starts beyond the order considered in this
paper, ${}_{-2}\eta_{\ell 2}=1+\mathcal{O}(\varepsilon_{\rm PM}^5)$.
Thus, the conservative information through $4$PM is fully contained in the real
phase shifts
\begin{align}
  {}_{-2}\delta^P_{\ell 2}
  =
  \sum_{n\geq1}
  \varepsilon_{\rm PM}^n\,
  {}_{-2}\delta^{P,(n)}_{\ell 2}\,.
\end{align}
From the matrix element
$\mathcal{N}_{\lambda_1\lambda_2}^{(n)}(x,\phi)$,  we define the helicity-preserving and helicity-reversing partial-wave modes from the matrix element
$\mathcal{N}_{\lambda_1\lambda_2}^{(n)}$
\begin{align}
  A_\ell^{(n)}
  &\equiv
  \frac{\omega}{4\pi\sqrt{\pi(2\ell+1)}}
  \int_{\Omega}\,
  {}_{2}Y_{\ell,-2}(\theta,\phi)\,
  \overline{\mathcal{N}}_{++}^{(n)}(x,\phi)\,,\label{eq:partial-wave-projections}\\
  B_\ell^{(n)}
  &\equiv
  \frac{\omega}{4\pi\sqrt{\pi(2\ell+1)}}
  \int_{\Omega}\,
  {}_{2}Y_{\ell,-2}(\pi-\theta,\phi)\,
  \overline{\mathcal{N}}_{+-}^{(n)}(x,\phi)\notag\,.
\end{align}
The overline denotes the standard partial-wave regularisation of the leading
helicity-preserving Coulomb pole. At the $4$PM order, no new forward singularity is
expected in the $\mathcal{N}$-matrix element.

The matching statement is that these two WQFT projections equal the two
independent parity combinations of the BHPT phase shift~\cite{Bautista:2026qse}
\begin{align}
  A_\ell^{(n)}
  &=
  \sum_{P=\pm1}
  {}_{-2}\delta^{P,(n)}_{\ell 2}\nonumber\,,
  \\
  B_\ell^{(n)}
  &=
  (-1)^\ell
  \sum_{P=\pm1}
  P\,{}_{-2}\delta^{P,(n)}_{\ell 2}\,.
  \label{eq:wqft-bhpt-matching}
\end{align}
At 4PM, we checked explicitly that Eq.~\eqref{eq:wqft-bhpt-matching} is numerically satisfied up to $\ell=10$, using publicly available codes for the computation of BHPT coefficients~\cite{BlackHolePerturbationToolkit,Markovic:2025kvr}.
\section{Outlook}
We have computed the minimal spinless gravitational Compton amplitude
at fourth post-Minkowskian order using WQFT. This provides, to our knowledge, the first fully analytic three-loop determination of the classical spinless Compton amplitude in this setup.
The calculation required the construction of the three-loop integrand, its reduction to 15 master integrals, and their subsequent evaluation. The master integrals were solved analytically using the method of canonical differential equations, revealing the presence of an elliptic sector.
After subtracting the lower-order Born iterations,
we extracted the corresponding fourth-order $N$-matrix element
and found agreement with the spinless BHPT phase shift.

Our result opens several directions.
First, the integral basis, reduction strategy, and analytic methods developed here provide the natural starting point for including spin and spin-induced finite-size effects.
Second, the same framework can be used to determine non-minimal worldline couplings, including tidal and absorptive response coefficients, by matching to black-hole perturbation theory at higher post-Minkowskian orders.
Finally, the emergence of elliptic sectors already at three loops indicates that higher-order gravitational-wave scattering probes increasingly rich classes of special functions. The present computation therefore provides both a concrete fourth-order result and a template for extending amplitude-based methods to more general compact-object dynamics.
\begin{acknowledgments}
We are very grateful to the authors of~\cite{Bautista:2026fcp} for sharing preliminary results for the $T$-matrix and for coordinating on submission.
We wish to thank Stefano De Angelis, Gustav Jakobsen, Pierpaolo Mastrolia, and Lorenzo Tancredi for useful discussions.

We would like to further acknowledge Xiao Liu and Yan-Qing Ma for their invaluable help in the use of \textsc{AMFlow}.
The computations for this work were performed with computing and storage resources provided by the CloudVeneto initiative at the University of Padova and INFN.

G.B.'s research is supported by the Italian MIUR under contract 20223ANFHR (PRIN2022), by the ERC (NOTIMEFORCOSMO, 101126304),
and by the INFN initiatives \textit{Amplitudes} and \textit{TPPC}.
S.S.'s research is partially supported by the INFN initiatives \textit{Amplitudes}.

\end{acknowledgments}

\bibliographystyle{apsrev4-1}
\bibliography{binary.bib}

@article{Bautista:2026qse,
    author = "Bautista, Yilber Fabian and Driesse, Mathias and Haddad, Kays and Jakobsen, Gustav Uhre",
    title = "{Gravitational wave scattering in spinless WQFT}",
    eprint = "2602.06125",
    archivePrefix = "arXiv",
    primaryClass = "hep-th",
    reportNumber = "HU-EP-26/05",
    doi = "10.1007/JHEP05(2026)252",
    journal = "JHEP",
    volume = "05",
    pages = "252",
    year = "2026"
}

@article{Mogull:2020sak,
    author = "Mogull, Gustav and Plefka, Jan and Steinhoff, Jan",
    title = "{Classical black hole scattering from a worldline quantum field theory}",
    eprint = "2010.02865",
    archivePrefix = "arXiv",
    primaryClass = "hep-th",
    reportNumber = "UUITP-37/20, HU-EP-20/22-RTG",
    doi = "10.1007/JHEP02(2021)048",
    journal = "JHEP",
    volume = "02",
    pages = "048",
    year = "2021"
}

@article{Bjerrum-Bohr:2025bqg,
    author = "Bjerrum-Bohr, N. Emil J. and Chen, Gang and Eriksen, Carl Jordan and Shah, Nabha",
    title = "{The gravitational Compton amplitude from flat and curved spacetimes at second post-Minkowskian order}",
    eprint = "2506.19705",
    archivePrefix = "arXiv",
    primaryClass = "hep-th",
    doi = "10.1007/JHEP10(2025)235",
    journal = "JHEP",
    volume = "10",
    pages = "235",
    year = "2025"
}

@article{Anastasiou:2023koq,
    author = "Anastasiou, Charalampos and Karlen, Julia and Vicini, Matilde",
    title = "{Tensor reduction of loop integrals}",
    eprint = "2308.14701",
    archivePrefix = "arXiv",
    primaryClass = "hep-ph",
    doi = "10.1007/JHEP12(2023)169",
    journal = "JHEP",
    volume = "12",
    pages = "169",
    year = "2023"
}

@article{Liu:2022chg,
    author = "Liu, Xiao and Ma, Yan-Qing",
    title = "{AMFlow: A Mathematica package for Feynman integrals computation via auxiliary mass flow}",
    eprint = "2201.11669",
    archivePrefix = "arXiv",
    primaryClass = "hep-ph",
    doi = "10.1016/j.cpc.2022.108565",
    journal = "Comput. Phys. Commun.",
    volume = "283",
    pages = "108565",
    year = "2023"
}

@article{Regge:1957td,
    author = "Regge, Tullio and Wheeler, John A.",
    title = "{Stability of a Schwarzschild singularity}",
    doi = "10.1103/PhysRev.108.1063",
    journal = "Phys. Rev.",
    volume = "108",
    pages = "1063--1069",
    year = "1957"
}

@article{Zerilli:1970se,
    author = "Zerilli, Frank J.",
    title = "{Effective Potential for Even-Parity Regge-Wheeler Gravitational Perturbation Equations}",
    doi = "10.1103/PhysRevLett.24.737",
    journal = "Phys. Rev. Lett.",
    volume = "24",
    pages = "737--738",
    year = "1970"
}

@article{Teukolsky:1973ha,
    author = "Teukolsky, Saul A.",
    title = "{Perturbations of a rotating black hole. 1. Fundamental equations for gravitational electromagnetic and neutrino field perturbations}",
    doi = "10.1086/152444",
    journal = "Astrophys. J.",
    volume = "185",
    pages = "635--647",
    year = "1973"
}

@article{Mano:1996vt,
    author = "Mano, Shuhei and Suzuki, Hisao and Takasugi, Eiichi",
    title = "{Analytic solutions of the Teukolsky equation and their low frequency expansions}",
    eprint = "gr-qc/9603020",
    archivePrefix = "arXiv",
    reportNumber = "OU-HET-238",
    doi = "10.1143/PTP.95.1079",
    journal = "Prog. Theor. Phys.",
    volume = "95",
    pages = "1079--1096",
    year = "1996"
}

@article{Duhr:2019tlz,
    author = "Duhr, Claude and Dulat, Falko",
    title = "{PolyLogTools {\textemdash} polylogs for the masses}",
    eprint = "1904.07279",
    archivePrefix = "arXiv",
    primaryClass = "hep-th",
    reportNumber = "CP3-19-17, CERN-TH-2019-045, SLAC-PUB-17423",
    doi = "10.1007/JHEP08(2019)135",
    journal = "JHEP",
    volume = "08",
    pages = "135",
    year = "2019"
}

@article{Dolan:2007mj,
    author = "Dolan, Sam R.",
    title = "{Instability of the massive Klein-Gordon field on the Kerr spacetime}",
    eprint = "0705.2880",
    archivePrefix = "arXiv",
    primaryClass = "gr-qc",
    doi = "10.1103/PhysRevD.76.084001",
    journal = "Phys. Rev. D",
    volume = "76",
    pages = "084001",
    year = "2007"
}

@article{Goldberger:2004jt,
    author = "Goldberger, Walter D. and Rothstein, Ira Z.",
    title = "{An Effective Field Theory of Gravity for Extended Objects}",
    eprint = "hep-th/0409156",
    archivePrefix = "arXiv",
    doi = "10.1103/PhysRevD.73.104029",
    journal = "Phys. Rev. D",
    volume = "73",
    pages = "104029",
    year = "2006"
}

@article{Bjerrum-Bohr:2026fhs,
    author = "Bjerrum-Bohr, N. Emil J. and Chen, Gang and Jordan Eriksen, Carl and Shah, Nabha",
    title = "{The gravitational Compton amplitude at third post-Minkowskian order}",
    eprint = "2602.06947",
    archivePrefix = "arXiv",
    primaryClass = "hep-th",
    reportNumber = "HU-EP-26/07-RTG",
    month = "2",
    year = "2026"
}

@article{Ivanov:2026icp,
    author = "Ivanov, Mikhail M. and Li, Yue-Zhou and Parra-Martinez, Julio and Zhou, Zihan",
    title = "{Gravitational Raman Scattering: a Systematic Toolkit for Tidal Effects in General Relativity}",
    eprint = "2602.06951",
    archivePrefix = "arXiv",
    primaryClass = "hep-th",
    reportNumber = "MIT-CTP/6001",
    month = "2",
    year = "2026"
}

@article{Ivanov:2022qqt,
    author = "Ivanov, Mikhail M. and Zhou, Zihan",
    title = "{Vanishing of Black Hole Tidal Love Numbers from Scattering Amplitudes}",
    eprint = "2209.14324",
    archivePrefix = "arXiv",
    primaryClass = "hep-th",
    doi = "10.1103/PhysRevLett.130.091403",
    journal = "Phys. Rev. Lett.",
    volume = "130",
    number = "9",
    pages = "091403",
    year = "2023"
}

@article{Saketh:2023bul,
    author = "Saketh, M. V. S. and Zhou, Zihan and Ivanov, Mikhail M.",
    title = "{Dynamical tidal response of Kerr black holes from scattering amplitudes}",
    eprint = "2307.10391",
    archivePrefix = "arXiv",
    primaryClass = "hep-th",
    doi = "10.1103/PhysRevD.109.064058",
    journal = "Phys. Rev. D",
    volume = "109",
    number = "6",
    pages = "064058",
    year = "2024"
}

@article{Ivanov:2024sds,
    author = "Ivanov, Mikhail M. and Li, Yue-Zhou and Parra-Martinez, Julio and Zhou, Zihan",
    title = "{Gravitational Raman Scattering in Effective Field Theory: A Scalar Tidal Matching at O(G3)}",
    eprint = "2401.08752",
    archivePrefix = "arXiv",
    primaryClass = "hep-th",
    reportNumber = "MIT-CTP/5664",
    doi = "10.1103/PhysRevLett.132.131401",
    journal = "Phys. Rev. Lett.",
    volume = "132",
    number = "13",
    pages = "131401",
    year = "2024",
    note = "[Erratum: Phys.Rev.Lett. 134, 159901 (2025)]"
}

@article{Caron-Huot:2025tlq,
    author = "Caron-Huot, Simon and Correia, Miguel and Isabella, Giulia and Solon, Mikhail",
    title = "{Gravitational Wave Scattering via the Born Series: Scalar Tidal Matching to O(G7) and Beyond}",
    eprint = "2503.13593",
    archivePrefix = "arXiv",
    primaryClass = "hep-th",
    doi = "10.1103/qd3c-nfz6",
    journal = "Phys. Rev. Lett.",
    volume = "135",
    number = "19",
    pages = "191601",
    year = "2025"
}

@article{Weinberg:1965nx,
    author = "Weinberg, Steven",
    title = "{Infrared photons and gravitons}",
    doi = "10.1103/PhysRev.140.B516",
    journal = "Phys. Rev.",
    volume = "140",
    pages = "B516--B524",
    year = "1965"
}

@article{Henn:2013pwa,
    author = "Henn, Johannes M.",
    title = "{Multiloop integrals in dimensional regularization made simple}",
    eprint = "1304.1806",
    archivePrefix = "arXiv",
    primaryClass = "hep-th",
    doi = "10.1103/PhysRevLett.110.251601",
    journal = "Phys. Rev. Lett.",
    volume = "110",
    pages = "251601",
    year = "2013"
}

@article{Meyer:2017joq,
    author = "Meyer, Christoph",
    title = "{Algorithmic transformation of multi-loop master integrals to a canonical basis with CANONICA}",
    eprint = "1705.06252",
    archivePrefix = "arXiv",
    primaryClass = "hep-ph",
    reportNumber = "HU-EP-17-10",
    doi = "10.1016/j.cpc.2017.09.014",
    journal = "Comput. Phys. Commun.",
    volume = "222",
    pages = "295--312",
    year = "2018"
}

@article{Cheung:2018wkq,
    author = "Cheung, Clifford and Rothstein, Ira Z. and Solon, Mikhail P.",
    title = "{From Scattering Amplitudes to Classical Potentials in the Post-Minkowskian Expansion}",
    eprint = "1808.02489",
    archivePrefix = "arXiv",
    primaryClass = "hep-th",
    reportNumber = "CALT-TH-2018-031",
    doi = "10.1103/PhysRevLett.121.251101",
    journal = "Phys. Rev. Lett.",
    volume = "121",
    number = "25",
    pages = "251101",
    year = "2018"
}

@article{Bern:2019nnu,
    author = "Bern, Zvi and Cheung, Clifford and Roiban, Radu and Shen, Chia-Hsien and Solon, Mikhail P. and Zeng, Mao",
    title = "{Scattering Amplitudes and the Conservative Hamiltonian for Binary Systems at Third Post-Minkowskian Order}",
    eprint = "1901.04424",
    archivePrefix = "arXiv",
    primaryClass = "hep-th",
    reportNumber = "CALT-TH 2019-002, UCLA/TEP/2019/101",
    doi = "10.1103/PhysRevLett.122.201603",
    journal = "Phys. Rev. Lett.",
    volume = "122",
    number = "20",
    pages = "201603",
    year = "2019"
}

@article{Bern:2021dqo,
    author = "Bern, Zvi and Parra-Martinez, Julio and Roiban, Radu and Ruf, Michael S. and Shen, Chia-Hsien and Solon, Mikhail P. and Zeng, Mao",
    title = "{Scattering Amplitudes and Conservative Binary Dynamics at ${\cal O}(G^4)$}",
    eprint = "2101.07254",
    archivePrefix = "arXiv",
    primaryClass = "hep-th",
    reportNumber = "CALT-TH-2021-004, FR-PHENO-2021-03, OUTP-21-03P",
    doi = "10.1103/PhysRevLett.126.171601",
    journal = "Phys. Rev. Lett.",
    volume = "126",
    number = "17",
    pages = "171601",
    year = "2021"
}

@article{Kosower:2018adc,
    author = "Kosower, David A. and Maybee, Ben and O'Connell, Donal",
    title = "{Amplitudes, Observables, and Classical Scattering}",
    eprint = "1811.10950",
    archivePrefix = "arXiv",
    primaryClass = "hep-th",
    doi = "10.1007/JHEP02(2019)137",
    journal = "JHEP",
    volume = "02",
    pages = "137",
    year = "2019"
}

@article{Damgaard:2021ipf,
    author = "Damgaard, Poul H. and Plante, Ludovic and Vanhove, Pierre",
    title = "{On an exponential representation of the gravitational S-matrix}",
    eprint = "2107.12891",
    archivePrefix = "arXiv",
    primaryClass = "hep-th",
    reportNumber = "IPhT-t21/037, CERN-TH-2021-111",
    doi = "10.1007/JHEP11(2021)213",
    journal = "JHEP",
    volume = "11",
    pages = "213",
    year = "2021"
}

@article{Jakobsen:2023ndj,
    author = "Jakobsen, Gustav Uhre and Mogull, Gustav and Plefka, Jan and Sauer, Benjamin and Xu, Yingxuan",
    title = "{Conservative Scattering of Spinning Black Holes at Fourth Post-Minkowskian Order}",
    eprint = "2306.01714",
    archivePrefix = "arXiv",
    primaryClass = "hep-th",
    reportNumber = "HU-EP-23/16-RTG",
    doi = "10.1103/PhysRevLett.131.151401",
    journal = "Phys. Rev. Lett.",
    volume = "131",
    number = "15",
    pages = "151401",
    year = "2023"
}

@article{Jakobsen:2022psy,
    author = "Jakobsen, Gustav Uhre and Mogull, Gustav and Plefka, Jan and Sauer, Benjamin",
    title = "{All things retarded: radiation-reaction in worldline quantum field theory}",
    eprint = "2207.00569",
    archivePrefix = "arXiv",
    primaryClass = "hep-th",
    reportNumber = "HU-EP-22/24-RTG",
    doi = "10.1007/JHEP10(2022)128",
    journal = "JHEP",
    volume = "10",
    pages = "128",
    year = "2022"
}

@article{Berends:1987me,
    author = "Berends, Frits A. and Giele, W. T.",
    title = "{Recursive Calculations for Processes with n Gluons}",
    reportNumber = "Print-88-0100 (LEIDEN)",
    doi = "10.1016/0550-3213(88)90442-7",
    journal = "Nucl. Phys. B",
    volume = "306",
    pages = "759--808",
    year = "1988"
}

@phdthesis{Jakobsen:2023oow,
    author = "Jakobsen, Gustav Uhre",
    title = "{Gravitational Scattering of Compact Bodies from Worldline Quantum Field Theory}",
    eprint = "2308.04388",
    archivePrefix = "arXiv",
    primaryClass = "hep-th",
    reportNumber = "HU-EP-23/45-RTG",
    doi = "10.18452/27075",
    school = "Humboldt U., Berlin, Humboldt U., Berlin (main)",
    year = "2023"
}

@article{Kalin:2022hph,
    author = {K{\"a}lin, Gregor and Neef, Jakob and Porto, Rafael A.},
    title = "{Radiation-reaction in the Effective Field Theory approach to Post-Minkowskian dynamics}",
    eprint = "2207.00580",
    archivePrefix = "arXiv",
    primaryClass = "hep-th",
    reportNumber = "DESY-22-109, DESY 22-109",
    doi = "10.1007/JHEP01(2023)140",
    journal = "JHEP",
    volume = "01",
    pages = "140",
    year = "2023"
}

@article{Schwinger:1960qe,
    author = "Schwinger, Julian S.",
    title = "{Brownian motion of a quantum oscillator}",
    doi = "10.1063/1.1703727",
    journal = "J. Math. Phys.",
    volume = "2",
    pages = "407--432",
    year = "1961"
}

@article{Keldysh:1964ud,
    author = "Keldysh, L. V.",
    title = "{Diagram Technique for Nonequilibrium Processes}",
    doi = "10.1142/9789811279461_0007",
    journal = "Sov. Phys. JETP",
    volume = "20",
    pages = "1018--1026",
    year = "1965"
}

@article{Gorges:2023zgv,
    author = {G{\"o}rges, Lennard and Nega, Christoph and Tancredi, Lorenzo and Wagner, Fabian J.},
    title = "{On a procedure to derive {\ensuremath{\epsilon}}-factorised differential equations beyond polylogarithms}",
    eprint = "2305.14090",
    archivePrefix = "arXiv",
    primaryClass = "hep-th",
    doi = "10.1007/JHEP07(2023)206",
    journal = "JHEP",
    volume = "07",
    pages = "206",
    year = "2023"
}

@article{Duhr:2025lbz,
    author = "Duhr, Claude and Maggio, Sara and Nega, Christoph and Sauer, Benjamin and Tancredi, Lorenzo and Wagner, Fabian J.",
    title = "{Aspects of canonical differential equations for Calabi-Yau geometries and beyond}",
    eprint = "2503.20655",
    archivePrefix = "arXiv",
    primaryClass = "hep-th",
    reportNumber = "BONN-TH-2025-11, TUM-HEP 1559/25, HU-EP-25/13-RTG",
    doi = "10.1007/JHEP06(2025)128",
    journal = "JHEP",
    volume = "06",
    pages = "128",
    year = "2025"
}

@article{Forner:2026vby,
    author = "Forner, Felix and Mella, Cesare Carlo and Nega, Christoph and Tancredi, Lorenzo and Wagner, Fabian J.",
    title = "{Integrand Analysis, Leading Singularities and Canonical Bases beyond Polylogarithms}",
    eprint = "2604.25270",
    archivePrefix = "arXiv",
    primaryClass = "hep-th",
    reportNumber = "OUTP-26-03P, TUM-HEP-1601/26",
    month = "4",
    year = "2026"
}

@article{Broedel:2018qkq,
    author = "Broedel, Johannes and Duhr, Claude and Dulat, Falko and Penante, Brenda and Tancredi, Lorenzo",
    title = "{Elliptic Feynman integrals and pure functions}",
    eprint = "1809.10698",
    archivePrefix = "arXiv",
    primaryClass = "hep-th",
    reportNumber = "CP3-18-58, CERN-TH-2018-211, HU-Mathematik-2018-09, HU-EP-18/29, SLAC-PUB-17336",
    doi = "10.1007/JHEP01(2019)023",
    journal = "JHEP",
    volume = "01",
    pages = "023",
    year = "2019"
}

@article{Beneke:1997zp,
    author = "Beneke, M. and Smirnov, Vladimir A.",
    title = "{Asymptotic expansion of Feynman integrals near threshold}",
    eprint = "hep-ph/9711391",
    archivePrefix = "arXiv",
    reportNumber = "CERN-TH-97-315",
    doi = "10.1016/S0550-3213(98)00138-2",
    journal = "Nucl. Phys. B",
    volume = "522",
    pages = "321--344",
    year = "1998"
}

@article{Jantzen:2012mw,
    author = "Jantzen, Bernd and Smirnov, Alexander V. and Smirnov, Vladimir A.",
    title = "{Expansion by regions: revealing potential and Glauber regions automatically}",
    eprint = "1206.0546",
    archivePrefix = "arXiv",
    primaryClass = "hep-ph",
    reportNumber = "TTK-12-21, TTP12-015, SFB-CPP-12-31",
    doi = "10.1140/epjc/s10052-012-2139-2",
    journal = "Eur. Phys. J. C",
    volume = "72",
    pages = "2139",
    year = "2012"
}

@article{Pak:2010pt,
    author = "Pak, A. and Smirnov, A.",
    title = "{Geometric approach to asymptotic expansion of Feynman integrals}",
    eprint = "1011.4863",
    archivePrefix = "arXiv",
    primaryClass = "hep-ph",
    doi = "10.1140/epjc/s10052-011-1626-1",
    journal = "Eur. Phys. J. C",
    volume = "71",
    pages = "1626",
    year = "2011"
}

@article{Mastrolia:2017pfy,
    author = "Mastrolia, Pierpaolo and Passera, Massimo and Primo, Amedeo and Schubert, Ulrich",
    title = "{Master integrals for the NNLO virtual corrections to $\mu e$ scattering in QED: the planar graphs}",
    eprint = "1709.07435",
    archivePrefix = "arXiv",
    primaryClass = "hep-ph",
    doi = "10.1007/JHEP11(2017)198",
    journal = "JHEP",
    volume = "11",
    pages = "198",
    year = "2017"
}

@article{Dulat:2014mda,
    author = "Dulat, Falko and Mistlberger, Bernhard",
    title = "{Real-Virtual-Virtual contributions to the inclusive Higgs cross section at N3LO}",
    eprint = "1411.3586",
    archivePrefix = "arXiv",
    primaryClass = "hep-ph",
    month = "11",
    year = "2014"
}

@article{Chestnov:2023kww,
    author = "Chestnov, Vsevolod and Matsubara-Heo, Saiei J. and Munch, Henrik J. and Takayama, Nobuki",
    title = "{Restrictions of Pfaffian systems for Feynman integrals}",
    eprint = "2305.01585",
    archivePrefix = "arXiv",
    primaryClass = "hep-th",
    doi = "10.1007/JHEP11(2023)202",
    journal = "JHEP",
    volume = "11",
    pages = "202",
    year = "2023"
}

@article{Brunello:2025cot,
    author = "Brunello, Giacomo and Chestnov, Vsevolod and Crisanti, Giulio and Giroux, Mathieu and Smith, Sid",
    title = "{Gravitational waveforms from restriction theory and rapid-decay homology}",
    eprint = "2510.26874",
    archivePrefix = "arXiv",
    primaryClass = "hep-th",
    doi = "10.1103/ynk9-ykw7",
    journal = "Phys. Rev. D",
    volume = "113",
    number = "8",
    pages = "085011",
    year = "2026"
}

@article{Chen:1977oja,
    author = "Chen, Kuo-Tsai",
    title = "{Iterated path integrals}",
    doi = "10.1090/S0002-9904-1977-14320-6",
    journal = "Bull. Am. Math. Soc.",
    volume = "83",
    pages = "831--879",
    year = "1977"
}

@article{Goldberger:2005cd,
    author = "Goldberger, Walter D. and Rothstein, Ira Z.",
    title = "{Dissipative effects in the worldline approach to black hole dynamics}",
    eprint = "hep-th/0511133",
    archivePrefix = "arXiv",
    doi = "10.1103/PhysRevD.73.104030",
    journal = "Phys. Rev. D",
    volume = "73",
    pages = "104030",
    year = "2006"
}

@article{Tkachov:1981wb,
    author = "Tkachov, F. V.",
    title = "{A theorem on analytical calculability of 4-loop renormalization group functions}",
    doi = "10.1016/0370-2693(81)90288-4",
    journal = "Phys. Lett. B",
    volume = "100",
    pages = "65--68",
    year = "1981"
}

@article{Chetyrkin:1981qh,
    author = "Chetyrkin, K. G. and Tkachov, F. V.",
    title = "{Integration by parts: The algorithm to calculate $\beta$-functions in 4 loops}",
    doi = "10.1016/0550-3213(81)90199-1",
    journal = "Nucl. Phys. B",
    volume = "192",
    pages = "159--204",
    year = "1981"
}

@article{Laporta:2000dsw,
    author = "Laporta, S.",
    title = "{High-precision calculation of multiloop Feynman integrals by difference equations}",
    eprint = "hep-ph/0102033",
    archivePrefix = "arXiv",
    doi = "10.1142/S0217751X00002159",
    journal = "Int. J. Mod. Phys. A",
    volume = "15",
    pages = "5087--5159",
    year = "2000"
}

@article{Remiddi:1997ny,
    author = "Remiddi, Ettore",
    title = "{Differential equations for Feynman graph amplitudes}",
    eprint = "hep-th/9711188",
    archivePrefix = "arXiv",
    reportNumber = "DFUB-97-15, DFUB 97-15",
    doi = "10.1007/BF03185566",
    journal = "Nuovo Cim. A",
    volume = "110",
    pages = "1435--1452",
    year = "1997"
}

@article{Kotikov:1990kg,
    author = "Kotikov, A. V.",
    title = "{Differential equations method: New technique for massive Feynman diagrams calculation}",
    reportNumber = "ITF-90-31E",
    doi = "10.1016/0370-2693(91)90413-K",
    journal = "Phys. Lett. B",
    volume = "254",
    pages = "158--164",
    year = "1991"
}

@article{Kotikov:1991pm,
    author = "Kotikov, A. V.",
    title = "{Differential equation method: The Calculation of $n$-point Feynman diagrams}",
    doi = "10.1016/0370-2693(91)90536-Y",
    journal = "Phys. Lett. B",
    number = "1",
    volume = "267",
    pages = "123--127",
    year = "1991",
    note = "[Erratum: Phys.Lett.B 295, 409--409 (1992)]"
}

@article{Gehrmann:1999as,
    author = "Gehrmann, T. and Remiddi, E.",
    title = "{Differential equations for two-loop four-point functions}",
    eprint = "hep-ph/9912329",
    archivePrefix = "arXiv",
    reportNumber = "TTP-99-49",
    doi = "10.1016/S0550-3213(00)00223-6",
    journal = "Nucl. Phys. B",
    volume = "580",
    pages = "485--518",
    year = "2000"
}

@article{Argeri:2014qva,
    author = "Argeri, Mario and Di Vita, Stefano and Mastrolia, Pierpaolo and Mirabella, Edoardo and Schlenk, Johannes and Schubert, Ulrich and Tancredi, Lorenzo",
    title = "{Magnus and Dyson Series for Master Integrals}",
    eprint = "1401.2979",
    archivePrefix = "arXiv",
    primaryClass = "hep-ph",
    doi = "10.1007/JHEP03(2014)082",
    journal = "JHEP",
    volume = "03",
    pages = "082",
    year = "2014"
}

@article{Argeri:2007up,
    author = "Argeri, Mario and Mastrolia, Pierpaolo",
    title = "{Feynman Diagrams and Differential Equations}",
    eprint = "0707.4037",
    archivePrefix = "arXiv",
    primaryClass = "hep-ph",
    reportNumber = "ZU-TH-19-07",
    doi = "10.1142/S0217751X07037147",
    journal = "Int. J. Mod. Phys. A",
    volume = "22",
    pages = "4375--4436",
    year = "2007"
}

@article{Foffa:2013qca,
  author        = {Foffa, Stefano and Sturani, Riccardo},
  title         = {{Effective field theory methods to model compact binaries}},
  eprint        = {1309.3474},
  archiveprefix = {arXiv},
  primaryclass  = {gr-qc},
  doi           = {10.1088/0264-9381/31/4/043001},
  journal       = {Class. Quant. Grav.},
  volume        = {31},
  number        = {4},
  pages         = {043001},
  year          = {2014}
}

@article{Foffa:2016rgu,
  author        = {Foffa, Stefano and Mastrolia, Pierpaolo and Sturani, Riccardo and Sturm, Christian},
  title         = {{Effective field theory approach to the gravitational two-body dynamics, at fourth post-Newtonian order and quintic in the Newton constant}},
  eprint        = {1612.00482},
  archiveprefix = {arXiv},
  primaryclass  = {gr-qc},
  doi           = {10.1103/PhysRevD.95.104009},
  journal       = {Phys. Rev. D},
  volume        = {95},
  number        = {10},
  pages         = {104009},
  year          = {2017}
}

@article{Foffa:2019hrb,
  author        = {Foffa, Stefano and Mastrolia, Pierpaolo and Sturani, Riccardo and Sturm, Christian and Torres Bobadilla, William J.},
  title         = {{Static two-body potential at fifth post-Newtonian order}},
  eprint        = {1902.10571},
  archiveprefix = {arXiv},
  primaryclass  = {gr-qc},
  doi           = {10.1103/PhysRevLett.122.241605},
  journal       = {Phys. Rev. Lett.},
  volume        = {122},
  number        = {24},
  pages         = {241605},
  year          = {2019}
}

@article{Foffa:2019rdf,
  author        = {Foffa, Stefano and Sturani, Riccardo},
  title         = {{Conservative dynamics of binary systems to fourth Post-Newtonian order in the EFT approach I: Regularized Lagrangian}},
  eprint        = {1903.05113},
  archiveprefix = {arXiv},
  primaryclass  = {gr-qc},
  doi           = {10.1103/PhysRevD.100.024047},
  journal       = {Phys. Rev. D},
  volume        = {100},
  number        = {2},
  pages         = {024047},
  year          = {2019}
}

@article{Foffa:2019yfl,
  author        = {Foffa, Stefano and Porto, Rafael A. and Rothstein, Ira and Sturani, Riccardo},
  title         = {{Conservative dynamics of binary systems to fourth Post-Newtonian order in the EFT approach II: Renormalized Lagrangian}},
  eprint        = {1903.05118},
  archiveprefix = {arXiv},
  primaryclass  = {gr-qc},
  doi           = {10.1103/PhysRevD.100.024048},
  journal       = {Phys. Rev. D},
  volume        = {100},
  number        = {2},
  pages         = {024048},
  year          = {2019}
}

@article{Porto:2026fsd,
    author = "Porto, Rafael A. and Riva, Massimiliano M.",
    title = "{Black Hole Dynamics at Fifth Post-Newtonian Order}",
    eprint = "2604.09545",
    archivePrefix = "arXiv",
    primaryClass = "gr-qc",
    reportNumber = "DESY 26-050",
    month = "4",
    year = "2026"
}

@article{Porto:2024cwd,
    author = "Porto, Rafael A. and Riva, Massimiliano M. and Yang, Zixin",
    title = "{Nonlinear gravitational radiation reaction: failed tail, memories {\&} squares}",
    eprint = "2409.05860",
    archivePrefix = "arXiv",
    primaryClass = "gr-qc",
    reportNumber = "DESY 24-133",
    doi = "10.1007/JHEP04(2025)050",
    journal = "JHEP",
    volume = "04",
    pages = "050",
    year = "2025"
}

@article{Brunello:2025gpf,
    author = "Brunello, Giacomo and Mandal, Manoj K. and Mastrolia, Pierpaolo and Patil, Raj and Pegorin, Matteo and Ronca, Jonathan and Smith, Sid and Steinhoff, Jan and Torres Bobadilla, William J.",
    title = "{Six-loop gravitational interactions at the sixth post-Newtonian order}",
    eprint = "2512.19498",
    archivePrefix = "arXiv",
    primaryClass = "hep-th",
    reportNumber = "HU-EP-25/43-RTG",
    month = "12",
    year = "2025"
}

@article{Brunello:2026anu,
    author = "Brunello, Giacomo and Mandal, Manoj K. and Mastrolia, Pierpaolo and Patil, Raj and Pegorin, Matteo and Smith, Sid and Steinhoff, Jan",
    title = "{All-order structure of static gravitational interactions and the seventh post-Newtonian potential}",
    eprint = "2604.14134",
    archivePrefix = "arXiv",
    primaryClass = "hep-th",
    reportNumber = "HU-EP-26/15-RTG",
    month = "4",
    year = "2026"
}

@article{Kalin:2020mvi,
    author = {K{\"a}lin, Gregor and Porto, Rafael A.},
    title = "{Post-Minkowskian Effective Field Theory for Conservative Binary Dynamics}",
    eprint = "2006.01184",
    archivePrefix = "arXiv",
    primaryClass = "hep-th",
    reportNumber = "DESY20-077, SLAC-PUB-17529",
    doi = "10.1007/JHEP11(2020)106",
    journal = "JHEP",
    volume = "11",
    pages = "106",
    year = "2020"
}

@article{Flieger:2022xyq,
    author = "Flieger, Wojciech and Torres Bobadilla, William J.",
    title = "{Landau and leading singularities in arbitrary space-time dimensions}",
    eprint = "2210.09872",
    archivePrefix = "arXiv",
    primaryClass = "hep-th",
    reportNumber = "MPP-2022-129",
    doi = "10.1140/epjp/s13360-024-05796-7",
    journal = "Eur. Phys. J. Plus",
    volume = "139",
    number = "11",
    pages = "1022",
    year = "2024"
}

@article{Driesse:2026qiz,
    author = "Driesse, Mathias and Jakobsen, Gustav Uhre and Mogull, Gustav and Nega, Christoph and Plefka, Jan and Sauer, Benjamin and Usovitsch, Johann",
    title = "{Conservative Black Hole Scattering at Fifth Post-Minkowskian and Second Self-Force Order}",
    eprint = "2601.16256",
    archivePrefix = "arXiv",
    primaryClass = "hep-th",
    reportNumber = "HU-EP-26/04-RTG",
    month = "1",
    year = "2026"
}

@article{Dlapa:2026oyq,
    author = {Dlapa, Christoph and K{\"a}lin, Gregor and Liu, Zhengwen and Porto, Rafael A.},
    title = "{Nonlocal-in-time tail effects in gravitational scattering to fifth Post-Minkowskian and tenth self-force orders}",
    eprint = "2604.25916",
    archivePrefix = "arXiv",
    primaryClass = "hep-th",
    reportNumber = "DESY 26-055",
    month = "4",
    year = "2026"
}

@article{Bern:2025wyd,
    author = "Bern, Zvi and Herrmann, Enrico and Roiban, Radu and Ruf, Michael S. and Smirnov, Alexander V. and Smith, Sid and Zeng, Mao",
    title = "{Scattering Amplitudes and Conservative Binary Dynamics at $O(G^5)$ without Self-Force Truncation}",
    eprint = "2512.23654",
    archivePrefix = "arXiv",
    primaryClass = "hep-th",
    month = "12",
    year = "2025"
}

@article{Brunello:2025eso,
    author = "Brunello, Giacomo and De Angelis, Stefano and Kosower, David A.",
    title = "{Analytic one-loop scattering waveform in general relativity}",
    eprint = "2511.05412",
    archivePrefix = "arXiv",
    primaryClass = "hep-th",
    doi = "10.1103/6xtl-qn8z",
    journal = "Phys. Rev. D",
    volume = "114",
    number = "4",
    pages = "045004",
    year = "2026"
}

@article{Bohnenblust:2025gir,
    author = "Bohnenblust, Lara and Ita, Harald and Kraus, Manfred and Schlenk, Johannes",
    title = "{Gravitational Bremsstrahlung in Black-Hole Scattering at $\mathcal{O}(G^3)$: Quadratic-in-Spin Effects}",
    eprint = "2505.15724",
    archivePrefix = "arXiv",
    primaryClass = "hep-th",
    doi = "10.1007/JHEP12(2025)100",
    journal = "JHEP",
    volume = "12",
    pages = "100",
    year = "2025"
}

@article{Bjerrum-Bohr:2018xdl,
  author        = {Bjerrum-Bohr, N. E. J. and Damgaard, Poul H. and Festuccia, Guido and Plant\'e, Ludovic and Vanhove, Pierre},
  title         = {{General Relativity from Scattering Amplitudes}},
  eprint        = {1806.04920},
  archiveprefix = {arXiv},
  primaryclass  = {hep-th},
  doi           = {10.1103/PhysRevLett.121.171601},
  journal       = {Phys. Rev. Lett.},
  volume        = {121},
  number        = {17},
  pages         = {171601},
  year          = {2018}
}

@article{Bern:2019crd,
  author        = {Bern, Zvi and Cheung, Clifford and Roiban, Radu and Shen, Chia-Hsien and Solon, Mikhail P. and Zeng, Mao},
  title         = {{Black Hole Binary Dynamics from the Double Copy and Effective Theory}},
  eprint        = {1908.01493},
  archiveprefix = {arXiv},
  primaryclass  = {hep-th},
  reportnumber  = {CERN-TH-2019-128, CALT-TH 2019-026, UCLA/TEP/2019/103},
  doi           = {10.1007/JHEP10(2019)206},
  journal       = {JHEP},
  volume        = {10},
  pages         = {206},
  year          = {2019}
}

@article{Bern:2022jvn,
  author  = {Bern, Zvi and Parra-Martinez, Julio and Roiban, Radu and Ruf, Michael S. and Shen, Chia-Hsien and Solon, Mikhail P. and Zeng, Mao},
  title   = {{Scattering amplitudes and conservative dynamics at the fourth post-Minkowskian order}},
  doi     = {10.22323/1.416.0051},
  journal = {PoS},
  volume  = {LL2022},
  pages   = {051},
  year    = {2022}
}

@article{Bern:2023ccb,
  author        = {Bern, Zvi and Herrmann, Enrico and Roiban, Radu and Ruf, Michael S. and Smirnov, Alexander V. and Smirnov, Vladimir A. and Zeng, Mao},
  title         = {{Conservative Binary Dynamics at Order \ensuremath{\alpha}5 in Electrodynamics}},
  eprint        = {2305.08981},
  archiveprefix = {arXiv},
  primaryclass  = {hep-th},
  doi           = {10.1103/PhysRevLett.132.251601},
  journal       = {Phys. Rev. Lett.},
  volume        = {132},
  number        = {25},
  pages         = {251601},
  year          = {2024}
}

@article{Bern:2024adl,
  author        = {Bern, Zvi and Herrmann, Enrico and Roiban, Radu and Ruf, Michael S. and Smirnov, Alexander V. and Smirnov, Vladimir A. and Zeng, Mao},
  title         = {{Amplitudes, supersymmetric black hole scattering at $ \mathcal{O}\left({G}^5\right) $, and loop integration}},
  eprint        = {2406.01554},
  archiveprefix = {arXiv},
  primaryclass  = {hep-th},
  doi           = {10.1007/JHEP10(2024)023},
  journal       = {JHEP},
  volume        = {10},
  pages         = {023},
  year          = {2024}
}

@article{Parra-Martinez:2020dzs,
  author        = {Parra-Martinez, Julio and Ruf, Michael S. and Zeng, Mao},
  title         = {{Extremal black hole scattering at $\mathcal{O}(G^3)$: graviton dominance, eikonal exponentiation, and differential equations}},
  eprint        = {2005.04236},
  archiveprefix = {arXiv},
  primaryclass  = {hep-th},
  reportnumber  = {FR-PHENO-2020-007, UCLA/TEP/2020/103},
  doi           = {10.1007/JHEP11(2020)023},
  journal       = {JHEP},
  volume        = {11},
  pages         = {023},
  year          = {2020}
}

@article{Cheung:2020gyp,
  author        = {Cheung, Clifford and Solon, Mikhail P.},
  title         = {{Classical gravitational scattering at $ \mathcal{O} $(G$^{3}$) from Feynman diagrams}},
  eprint        = {2003.08351},
  archiveprefix = {arXiv},
  primaryclass  = {hep-th},
  reportnumber  = {CALT-TH-2020-006},
  doi           = {10.1007/JHEP06(2020)144},
  journal       = {JHEP},
  volume        = {06},
  pages         = {144},
  year          = {2020}
}

@article{DiVecchia:2021bdo,
  author        = {Di Vecchia, Paolo and Heissenberg, Carlo and Russo, Rodolfo and Veneziano, Gabriele},
  title         = {{The eikonal approach to gravitational scattering and radiation at $ \mathcal{O} $(G$^{3}$)}},
  eprint        = {2104.03256},
  archiveprefix = {arXiv},
  primaryclass  = {hep-th},
  reportnumber  = {CERN-TH-2021-046, NORDITA 2021-028, QMUL-PH-21-17},
  doi           = {10.1007/JHEP07(2021)169},
  journal       = {JHEP},
  volume        = {07},
  pages         = {169},
  year          = {2021}
}

@article{Brandhuber:2021eyq,
  author        = {Brandhuber, Andreas and Chen, Gang and Travaglini, Gabriele and Wen, Congkao},
  title         = {{Classical gravitational scattering from a gauge-invariant double copy}},
  eprint        = {2108.04216},
  archiveprefix = {arXiv},
  primaryclass  = {hep-th},
  reportnumber  = {QMUL-PH-21-18, SAGEX-21-07},
  doi           = {10.1007/JHEP10(2021)118},
  journal       = {JHEP},
  volume        = {10},
  pages         = {118},
  year          = {2021}
}

@article{Kalin:2020fhe,
  author        = {K\"alin, Gregor and Liu, Zhengwen and Porto, Rafael A.},
  title         = {{Conservative Dynamics of Binary Systems to Third Post-Minkowskian Order from the Effective Field Theory Approach}},
  eprint        = {2007.04977},
  archiveprefix = {arXiv},
  primaryclass  = {hep-th},
  reportnumber  = {DESY 20-114, DESY-20-114, SLAC-PUB-17545},
  doi           = {10.1103/PhysRevLett.125.261103},
  journal       = {Phys. Rev. Lett.},
  volume        = {125},
  number        = {26},
  pages         = {261103},
  year          = {2020}
}

@article{Jakobsen:2021smu,
  archiveprefix = {arXiv},
  author        = {Jakobsen, Gustav Uhre and Mogull, Gustav and Plefka, Jan and Steinhoff, Jan},
  doi           = {10.1103/PhysRevLett.126.201103},
  eprint        = {2101.12688},
  journal       = {Phys. Rev. Lett.},
  number        = {20},
  pages         = {201103},
  primaryclass  = {gr-qc},
  reportnumber  = {HU-EP-21/03-RTG},
  title         = {{Classical Gravitational Bremsstrahlung from a Worldline Quantum Field Theory}},
  volume        = {126},
  year          = {2021}
}

@article{Jakobsen:2023hig,
  author        = {Jakobsen, Gustav Uhre and Mogull, Gustav and Plefka, Jan and Sauer, Benjamin},
  title         = {{Dissipative Scattering of Spinning Black Holes at Fourth Post-Minkowskian Order}},
  eprint        = {2308.11514},
  archiveprefix = {arXiv},
  primaryclass  = {hep-th},
  reportnumber  = {HU-EP-23/47-RTG},
  doi           = {10.1103/PhysRevLett.131.241402},
  journal       = {Phys. Rev. Lett.},
  volume        = {131},
  number        = {24},
  pages         = {241402},
  year          = {2023}
}

@article{Mougiakakos:2021ckm,
  author        = {Mougiakakos, Stavros and Riva, Massimiliano Maria and Vernizzi, Filippo},
  title         = {{Gravitational Bremsstrahlung in the post-Minkowskian effective field theory}},
  eprint        = {2102.08339},
  archiveprefix = {arXiv},
  primaryclass  = {gr-qc},
  doi           = {10.1103/PhysRevD.104.024041},
  journal       = {Phys. Rev. D},
  volume        = {104},
  number        = {2},
  pages         = {024041},
  year          = {2021}
}

@article{Dlapa:2021npj,
  author        = {Dlapa, Christoph and K\"alin, Gregor and Liu, Zhengwen and Porto, Rafael A.},
  title         = {{Dynamics of binary systems to fourth Post-Minkowskian order from the effective field theory approach}},
  eprint        = {2106.08276},
  archiveprefix = {arXiv},
  primaryclass  = {hep-th},
  reportnumber  = {DESY 21-093, DESY-21-093, MPP-2021-83},
  doi           = {10.1016/j.physletb.2022.137203},
  journal       = {Phys. Lett. B},
  volume        = {831},
  pages         = {137203},
  year          = {2022}
}

@article{Dlapa:2022lmu,
  author        = {Dlapa, Christoph and K\"alin, Gregor and Liu, Zhengwen and Neef, Jakob and Porto, Rafael A.},
  title         = {{Radiation Reaction and Gravitational Waves at Fourth Post-Minkowskian Order}},
  eprint        = {2210.05541},
  archiveprefix = {arXiv},
  primaryclass  = {hep-th},
  doi           = {10.1103/PhysRevLett.130.101401},
  journal       = {Phys. Rev. Lett.},
  volume        = {130},
  number        = {10},
  pages         = {101401},
  year          = {2023}
}

@article{Damgaard:2023ttc,
  author        = {Damgaard, Poul H. and Hansen, Elias Roos and Plant\'e, Ludovic and Vanhove, Pierre},
  title         = {{Classical observables from the exponential representation of the gravitational S-matrix}},
  eprint        = {2307.04746},
  archiveprefix = {arXiv},
  primaryclass  = {hep-th},
  reportnumber  = {CERN-TH-2023-135, IPhT-T23/041, LAPTh-029/23},
  doi           = {10.1007/JHEP09(2023)183},
  journal       = {JHEP},
  volume        = {09},
  pages         = {183},
  year          = {2023}
}

@article{Driesse:2024feo,
    author = "Driesse, Mathias and Jakobsen, Gustav Uhre and Klemm, Albrecht and Mogull, Gustav and Nega, Christoph and Plefka, Jan and Sauer, Benjamin and Usovitsch, Johann",
    title = "{Emergence of Calabi{\textendash}Yau manifolds in high-precision black-hole scattering}",
    eprint = "2411.11846",
    archivePrefix = "arXiv",
    primaryClass = "hep-th",
    reportNumber = "HU-EP-24/32-RTG, QMUL-PH-24-26, BONN-TH-2024-15, TUM-HEP-1532/24",
    doi = "10.1038/s41586-025-08984-2",
    journal = "Nature",
    volume = "641",
    number = "8063",
    pages = "603--607",
    year = "2025"
}

@article{Driesse:2024xad,
  author        = {Driesse, Mathias and Jakobsen, Gustav Uhre and Mogull, Gustav and Plefka, Jan and Sauer, Benjamin and Usovitsch, Johann},
  title         = {{Conservative Black Hole Scattering at Fifth Post-Minkowskian and First Self-Force Order}},
  eprint        = {2403.07781},
  archiveprefix = {arXiv},
  primaryclass  = {hep-th},
  reportnumber  = {HU-EP-24/08-RTG, CERN-TH-2024-032},
  doi           = {10.1103/PhysRevLett.132.241402},
  journal       = {Phys. Rev. Lett.},
  volume        = {132},
  number        = {24},
  pages         = {241402},
  year          = {2024}
}

@article{Jakobsen:2021lvp,
  author        = {Jakobsen, Gustav Uhre and Mogull, Gustav and Plefka, Jan and Steinhoff, Jan},
  title         = {{Gravitational Bremsstrahlung and Hidden Supersymmetry of Spinning Bodies}},
  eprint        = {2106.10256},
  archiveprefix = {arXiv},
  primaryclass  = {hep-th},
  reportnumber  = {HU-EP-21/15-RTG},
  doi           = {10.1103/PhysRevLett.128.011101},
  journal       = {Phys. Rev. Lett.},
  volume        = {128},
  number        = {1},
  pages         = {011101},
  year          = {2022}
}

@article{DeAngelis:2023lvf,
  author        = {De Angelis, Stefano and Novichkov, Pavel P. and Gonzo, Riccardo},
  title         = {{Spinning waveforms from the Kosower-Maybee-O\textquoteright{}Connell formalism at leading order}},
  eprint        = {2309.17429},
  archiveprefix = {arXiv},
  primaryclass  = {hep-th},
  doi           = {10.1103/PhysRevD.110.L041502},
  journal       = {Phys. Rev. D},
  volume        = {110},
  number        = {4},
  pages         = {L041502},
  year          = {2024}
}

@article{Brandhuber:2023hhl,
  author        = {Brandhuber, Andreas and Brown, Graham R. and Chen, Gang and Gowdy, Joshua and Travaglini, Gabriele},
  title         = {{Resummed spinning waveforms from five-point amplitudes}},
  eprint        = {2310.04405},
  archiveprefix = {arXiv},
  primaryclass  = {hep-th},
  reportnumber  = {QMUL-PH-23-18},
  doi           = {10.1007/JHEP02(2024)026},
  journal       = {JHEP},
  volume        = {02},
  pages         = {026},
  year          = {2024}
}

@article{Aoude:2023dui,
  author        = {Aoude, Rafael and Haddad, Kays and Heissenberg, Carlo and Helset, Andreas},
  title         = {{Leading-order gravitational radiation to all spin orders}},
  eprint        = {2310.05832},
  archiveprefix = {arXiv},
  primaryclass  = {hep-th},
  doi           = {10.1103/PhysRevD.109.036007},
  journal       = {Phys. Rev. D},
  volume        = {109},
  number        = {3},
  pages         = {036007},
  year          = {2024}
}

@article{Brandhuber:2023hhy,
  archiveprefix = {arXiv},
  author        = {Brandhuber, Andreas and Brown, Graham R. and Chen, Gang and De Angelis, Stefano and Gowdy, Joshua and Travaglini, Gabriele},
  doi           = {10.1007/JHEP06(2023)048},
  eprint        = {2303.06111},
  journal       = {JHEP},
  pages         = {048},
  primaryclass  = {hep-th},
  reportnumber  = {QMUL-22-28,SAGEX-22-32-E},
  title         = {{One-loop gravitational bremsstrahlung and waveforms from a heavy-mass effective field theory}},
  volume        = {06},
  year          = {2023}
}

@article{Herderschee:2023fxh,
  archiveprefix = {arXiv},
  author        = {Herderschee, Aidan and Roiban, Radu and Teng, Fei},
  doi           = {10.1007/JHEP06(2023)004},
  eprint        = {2303.06112},
  journal       = {JHEP},
  pages         = {004},
  primaryclass  = {hep-th},
  reportnumber  = {LCTP-23-04},
  title         = {{The sub-leading scattering waveform from amplitudes}},
  volume        = {06},
  year          = {2023}
}

@article{Georgoudis:2023lgf,
  author        = {Georgoudis, Alessandro and Heissenberg, Carlo and Vazquez-Holm, Ingrid},
  title         = {{Inelastic exponentiation and classical gravitational scattering at one loop}},
  eprint        = {2303.07006},
  archiveprefix = {arXiv},
  primaryclass  = {hep-th},
  reportnumber  = {NORDITA 2023-010, UUITP-03/23},
  doi           = {10.1007/JHEP06(2023)126},
  journal       = {JHEP},
  volume        = {06},
  pages         = {126},
  year          = {2023}
}

@article{Elkhidir:2023dco,
    author = "Elkhidir, Asaad and O'Connell, Donal and Sergola, Matteo and Vazquez-Holm, Ingrid A.",
    title = "{Radiation and reaction at one loop}",
    eprint = "2303.06211",
    archivePrefix = "arXiv",
    primaryClass = "hep-th",
    doi = "10.1007/JHEP07(2024)272",
    journal = "JHEP",
    volume = "07",
    pages = "272",
    year = "2024"
}

@article{Caron-Huot:2023vxl,
  archiveprefix = {arXiv},
  author        = {Caron-Huot, Simon and Giroux, Mathieu and Hannesdottir, Holmfridur S. and Mizera, Sebastian},
  doi           = {10.1007/JHEP01(2024)139},
  eprint        = {2308.02125},
  journal       = {JHEP},
  pages         = {139},
  primaryclass  = {hep-th},
  title         = {{What can be measured asymptotically?}},
  volume        = {01},
  year          = {2024}
}

@article{Bohnenblust:2023qmy,
    author = "Bohnenblust, Lara and Ita, Harald and Kraus, Manfred and Schlenk, Johannes",
    title = "{Gravitational Bremsstrahlung in black-hole scattering at $ \mathcal{O}\left({G}^3\right) $: linear-in-spin effects}",
    eprint = "2312.14859",
    archivePrefix = "arXiv",
    primaryClass = "hep-th",
    reportNumber = "PSI-PR-23-47",
    doi = "10.1007/JHEP11(2024)109",
    journal = "JHEP",
    volume = "11",
    pages = "109",
    year = "2024"
}

@article{Akpinar:2025byi,
    author = "Akpinar, Dogan",
    title = "{Scattering gravitons off general spinning compact objects to O(G2S4)}",
    eprint = "2511.10280",
    archivePrefix = "arXiv",
    primaryClass = "hep-th",
    doi = "10.1103/1zs7-kj4f",
    journal = "Phys. Rev. D",
    volume = "113",
    number = "4",
    pages = "045003",
    year = "2026"
}

@article{Kim:2025gis,
    author = "Kim, Jung-Wook and Patil, Raj and Scheopner, Trevor and Steinhoff, Jan",
    title = "{Magnusian: relating the eikonal phase, the on-shell action, and the scattering generator}",
    eprint = "2511.05649",
    archivePrefix = "arXiv",
    primaryClass = "hep-th",
    reportNumber = "CERN-TH-2025-226, HU-EP-25/37-RTG",
    doi = "10.1007/JHEP03(2026)241",
    journal = "JHEP",
    volume = "03",
    pages = "241",
    year = "2026"
}

@article{DiVecchia:2023frv,
    author = "Di Vecchia, Paolo and Heissenberg, Carlo and Russo, Rodolfo and Veneziano, Gabriele",
    title = "{The gravitational eikonal: From particle, string and brane collisions to black-hole encounters}",
    eprint = "2306.16488",
    archivePrefix = "arXiv",
    primaryClass = "hep-th",
    reportNumber = "CERN-TH-2023-108, NORDITA 2023-026, QMUL-PH-23-09, UUITP-14/23",
    doi = "10.1016/j.physrep.2024.06.002",
    journal = "Phys. Rept.",
    volume = "1083",
    pages = "1--169",
    year = "2024"
}

@article{PRISM,
    author = "Brunello, Giacomo and Mastrolia, Pierpaolo and Ronca, Jonathan and Smith, Sid and Zeng, Mao",
    title = "{PRISM: Prime-field Reduction of Integrals using Syzygy Methods}",
    journal = "in progress"
}

@article{Gluza:2010ws,
    author = "Gluza, Janusz and Kajda, Krzysztof and Kosower, David A.",
    title = "{Towards a Basis for Planar Two-Loop Integrals}",
    eprint = "1009.0472",
    archivePrefix = "arXiv",
    primaryClass = "hep-th",
    reportNumber = "SACLAY-IPHT-T10-089, WIS-09-10-JULY-DPPA",
    doi = "10.1103/PhysRevD.83.045012",
    journal = "Phys. Rev. D",
    volume = "83",
    pages = "045012",
    year = "2011"
}

@article{Wu:2023upw,
    author = "Wu, Zihao and Boehm, Janko and Ma, Rourou and Xu, Hefeng and Zhang, Yang",
    title = "{NeatIBP 1.0, a package generating small-size integration-by-parts relations for Feynman integrals}",
    eprint = "2305.08783",
    archivePrefix = "arXiv",
    primaryClass = "hep-ph",
    reportNumber = "USTC-ICTS/PCFT-23-15",
    doi = "10.1016/j.cpc.2023.108999",
    journal = "Comput. Phys. Commun.",
    volume = "295",
    pages = "108999",
    year = "2024"
}

@article{Wu:2025aeg,
    author = {Wu, Zihao and B{\"o}hm, Janko and Ma, Rourou and Usovitsch, Johann and Xu, Yingxuan and Zhang, Yang},
    title = "{Performing integration-by-parts reductions using NeatIBP 1.1 + Kira}",
    eprint = "2502.20778",
    archivePrefix = "arXiv",
    primaryClass = "hep-ph",
    reportNumber = "USTC-ICTS/PCFT-25-10, MPP-2025-29, HU-EP-25/12-RTG",
    doi = "10.1016/j.cpc.2025.109798",
    journal = "Comput. Phys. Commun.",
    volume = "316",
    pages = "109798",
    year = "2025"
}

@article{Smith:2025xes,
    author = "Smith, Sid and Zeng, Mao",
    title = "{Feynman Integral Reduction using Syzygy-Constrained Symbolic Reduction Rules}",
    eprint = "2507.11140",
    archivePrefix = "arXiv",
    primaryClass = "hep-th",
    month = "7",
    year = "2025"
}

@article{Lange:2025fba,
    author = "Lange, Fabian and Usovitsch, Johann and Wu, Zihao",
    title = "{Kira 3: integral reduction with efficient seeding and optimized equation selection}",
    eprint = "2505.20197",
    archivePrefix = "arXiv",
    primaryClass = "hep-ph",
    reportNumber = "ZU-TH 39/25, HU-EP-25/17-RTG",
    month = "5",
    year = "2025"
}

@article{Peraro:2019svx,
    author = "Peraro, Tiziano",
    title = "{$\text{FiniteFlow}$: multivariate functional reconstruction using finite fields and dataflow graphs}",
    eprint = "1905.08019",
    archivePrefix = "arXiv",
    primaryClass = "hep-ph",
    reportNumber = "ZU-TH 24/19",
    doi = "10.1007/JHEP07(2019)031",
    journal = "JHEP",
    volume = "07",
    pages = "031",
    year = "2019"
}

@article{Remiddi:1999ew,
    author = "Remiddi, E. and Vermaseren, J. A. M.",
    title = "{Harmonic polylogarithms}",
    eprint = "hep-ph/9905237",
    archivePrefix = "arXiv",
    reportNumber = "NIKHEF-99-005, TTP-99-08",
    doi = "10.1142/S0217751X00000367",
    journal = "Int. J. Mod. Phys. A",
    volume = "15",
    pages = "725--754",
    year = "2000"
}

@article{Dlapa:2023hsl,
    author = {Dlapa, Christoph and K{\"a}lin, Gregor and Liu, Zhengwen and Porto, Rafael A.},
    title = "{Bootstrapping the relativistic two-body problem}",
    eprint = "2304.01275",
    archivePrefix = "arXiv",
    primaryClass = "hep-th",
    reportNumber = "DESY 23-041",
    doi = "10.1007/JHEP08(2023)109",
    journal = "JHEP",
    volume = "08",
    pages = "109",
    year = "2023"
}

@article{Laporta:2001rc,
    author = "Laporta, S.",
    title = "{High precision epsilon expansions of three loop master integrals contributing to the electron g-2 in QED}",
    eprint = "hep-ph/0111123",
    archivePrefix = "arXiv",
    doi = "10.1016/S0370-2693(01)01331-4",
    journal = "Phys. Lett. B",
    volume = "523",
    pages = "95--101",
    year = "2001"
}

@article{Lee:2010ik,
    author = "Lee, R. N. and Smirnov, V. A.",
    title = "{Analytic Epsilon Expansions of Master Integrals Corresponding to Massless Three-Loop Form Factors and Three-Loop g-2 up to Four-Loop Transcendentality Weight}",
    eprint = "1010.1334",
    archivePrefix = "arXiv",
    primaryClass = "hep-ph",
    doi = "10.1007/JHEP02(2011)102",
    journal = "JHEP",
    volume = "02",
    pages = "102",
    year = "2011"
}

@article{Chetyrkin:2003vi,
    author = "Chetyrkin, K. G. and Grozin, A. G.",
    title = "{Three loop anomalous dimension of the heavy light quark current in HQET}",
    eprint = "hep-ph/0303113",
    archivePrefix = "arXiv",
    reportNumber = "TTP03-10, SFB-CPP-03-05, THEP-03-05",
    doi = "10.1016/S0550-3213(03)00490-5",
    journal = "Nucl. Phys. B",
    volume = "666",
    pages = "289--302",
    year = "2003"
}

@article{Brandhuber:2025igz,
    author = "Brandhuber, Andreas and Brown, Graham R. and Pichini, Paolo and Travaglini, Gabriele and Vives Matasan, Pablo",
    title = "{The Magnus expansion in relativistic quantum field theory}",
    eprint = "2512.05017",
    archivePrefix = "arXiv",
    primaryClass = "hep-th",
    reportNumber = "QMUL-PH-25-27",
    month = "12",
    year = "2025"
}

@article{Jinno:2022sbr,
    author = {Jinno, Ryusuke and K{\"a}lin, Gregor and Liu, Zhengwen and Rubira, Henrique},
    title = "{Machine learning Post-Minkowskian integrals}",
    eprint = "2209.01091",
    archivePrefix = "arXiv",
    primaryClass = "hep-th",
    reportNumber = "IFT-UAM/CSIC-22-97, DESY-22-144, TUM-HEP 1392/22",
    doi = "10.1007/JHEP07(2023)181",
    journal = "JHEP",
    volume = "07",
    pages = "181",
    year = "2023"
}

@article{Falkowski:2024bgb,
    author = "Falkowski, Adam and Marinellis, Panagiotis",
    title = "{Spinning waveforms of scalar radiation in quadratic modified gravity}",
    eprint = "2407.16457",
    archivePrefix = "arXiv",
    primaryClass = "hep-th",
    doi = "10.1140/epjc/s10052-025-13814-w",
    journal = "Eur. Phys. J. C",
    volume = "85",
    number = "1",
    pages = "74",
    year = "2025"
}

@article{Brunello:2024ibk,
    author = "Brunello, Giacomo and De Angelis, Stefano",
    title = "{An improved framework for computing waveforms}",
    eprint = "2403.08009",
    archivePrefix = "arXiv",
    primaryClass = "hep-th",
    doi = "10.1007/JHEP07(2024)062",
    journal = "JHEP",
    volume = "07",
    pages = "062",
    year = "2024"
}

@book{Smirnov:2012gma,
    author = "Smirnov, Vladimir A.",
    title = "{Analytic tools for Feynman integrals}",
    doi = "10.1007/978-3-642-34886-0",
    volume = "250",
    year = "2012"
}

@article{Davies:2026cci,
    author = "Davies, J. and Kaneko, T. and Marinissen, C. and Ueda, T. and Vermaseren, J. A. M.",
    title = "{FORM Version 5.0}",
    eprint = "2601.19982",
    archivePrefix = "arXiv",
    primaryClass = "hep-ph",
    month = "1",
    year = "2026"
}

@article{Anastasiou:2002yz,
    author = "Anastasiou, Charalampos and Melnikov, Kirill",
    title = "{Higgs boson production at hadron colliders in NNLO QCD}",
    eprint = "hep-ph/0207004",
    archivePrefix = "arXiv",
    reportNumber = "SLAC-PUB-9273",
    doi = "10.1016/S0550-3213(02)00837-4",
    journal = "Nucl. Phys. B",
    volume = "646",
    pages = "220--256",
    year = "2002"
}

@article{Anastasiou:2003gr,
    author = "Anastasiou, Charalampos and Melnikov, Kirill and Petriello, Frank",
    title = "{A new method for real radiation at NNLO}",
    eprint = "hep-ph/0311311",
    archivePrefix = "arXiv",
    reportNumber = "SLAC-PUB-10252, UH-511-1039-03",
    doi = "10.1103/PhysRevD.69.076010",
    journal = "Phys. Rev. D",
    volume = "69",
    pages = "076010",
    year = "2004"
}

@misc{BlackHolePerturbationToolkit,
  title        = {{Black Hole Perturbation Toolkit}},
  howpublished = {\url{https://bhptoolkit.org}}
}

@article{Markovic:2025kvr,
    author = "Markovic, Jovan and Ivanov, Mikhail M.",
    title = "{A-BHPT-toolkit: Analytic Black Hole Perturbation Theory Package for Gravitational Scattering Amplitudes}",
    eprint = "2511.04765",
    archivePrefix = "arXiv",
    primaryClass = "gr-qc",
    reportNumber = "MIT-CTP/5954",
    month = "11",
    year = "2025"
}

@book{Weinberg:1995mt,
    author = "Weinberg, Steven",
    title = "{The Quantum theory of fields. Vol. 1: Foundations}",
    doi = "10.1017/CBO9781139644167",
    isbn = "978-0-521-67053-1, 978-0-511-25204-4",
    publisher = "Cambridge University Press",
    month = "6",
    year = "2005"
}

@book{Weinberg:1996kr,
    author = "Weinberg, Steven",
    title = "{The quantum theory of fields. Vol. 2: Modern applications}",
    doi = "10.1017/CBO9781139644174",
    isbn = "978-1-139-63247-8, 978-0-521-67054-8, 978-0-521-55002-4",
    publisher = "Cambridge University Press",
    month = "8",
    year = "2013"
}

@article{Correia:2024jgr,
    author = "Correia, Miguel and Isabella, Giulia",
    title = "{The Born regime of gravitational amplitudes}",
    eprint = "2406.13737",
    archivePrefix = "arXiv",
    primaryClass = "hep-th",
    doi = "10.1007/JHEP03(2025)144",
    journal = "JHEP",
    volume = "03",
    pages = "144",
    year = "2025"
}

@article{Bautista:2026fcp,
    author = "Bautista, Yilber Fabian and Driesse, Mathias and Haddad, Kays and Jakobsen, Gustav Uhre",
    title = "{Gravitational wave scattering at $\mathcal{O}(G^4)$: Murua construction and elliptics}",
    eprint = "2606.27544",
    archivePrefix = "arXiv",
    primaryClass = "hep-th",
    reportNumber = "HU-EP-26/19",
    month = "6",
    year = "2026"
}

\onecolumngrid
\newpage
\appendix
\section{Canonical Form}
\label{app:canonical}
The following set of integrals satisfies a system of canonical differential equations:
\begin{align}
    \begin{alignedat}{2}
        J_1 &= \frac{1}{\eps^{3}}F_{222000000000} \,,&\qquad
        J_2 &= \frac{(1-4\epsilon)(1-6\epsilon)}{15\epsilon^{2}}F_{100000001110}\,, 
        \\
        J_3 &= \frac{(1-2\epsilon)(1-   6\epsilon)}{5\epsilon^{2}}F_{101000000110}-2J_{1}-2J_{2}\,,& \qquad
        J_4 &= -\frac{(1-6\epsilon)}{5\epsilon^2}F_{210000001010}-J_1-J_2-J_3\,, 
        \\
        J_5 &= \frac{(1-4\epsilon)(1-6\epsilon)}{15\epsilon^2x^3}F_{00010000111-1}\,,& \qquad
        J_6 &= -\frac{(1-6\epsilon)}{5\epsilon}F_{001100001110}\,, 
        \\
        J_7 &= \frac{1}{w_0(x)}F_{010100001110}\,, &\qquad
        J_8 &= \frac{w_0(x)^2(x^3-x)}{\epsilon}\partial_x\!\left[\frac{1}{w_0(x)}F_{010100001110}\right]
        \\
        &&&+(2x^2-3)w_0(x)^2J_7\,,
        \\
        J_9 &= x\,F_{011100001110}\,,& \qquad
        J_{10} &= \frac{x(x^2-1)}{2\epsilon}\partial_x F_{011100001110}
        \\
        &&&+\frac{(1+x)\big[(1+2\epsilon)x-(1+6\epsilon)\big]}{2x\epsilon}J_{9}\,, 
        \\
        J_{11} &= -\frac{x}{5}F_{101100001110} \,, &\qquad
        J_{12} &= -\frac{(1-6\epsilon)}{5\epsilon}F_{10110-1001110}+\frac{4(1-2\epsilon)x}{\epsilon}J_{11}\,, 
        \\
        J_{13} &= x^2F_{111100001110}\,,& \qquad
        J_{14} &= F_{11110-1001110}\,, 
        \\
        J_{15} &= F_{1111-10001110}\,.& \qquad & \\
    \end{alignedat}
\end{align}  
Here \(w_0(x)\) denotes the first elliptic period. It satisfies the second-order differential equation:
\begin{align}
w_0''(x)
-
\frac{3x^2-1}{x(1-x^2)}\,w_0'(x)
-
\frac{1}{1-x^2}\,w_0(x)
=0 \, .
\end{align}
This system is in canonical form:
\begin{align}
    \partial_x \mathbf{J} = \epsilon A(x)\,  \mathbf{J} \, , \qquad  A(x)=\frac{A_{1}}{x+1}+\frac{A_{-1}}{x-1}+\frac{A_0}{x}+A_{w_0}(x) \, ,
\end{align}
where:
\begin{align}
    A_{1} = \left[\begin{smallmatrix}
        \mzero & \mzero & \mzero & \mzero & \mzero & \mzero & \mzero & \mzero & \mzero & \mzero & \mzero & \mzero & \mzero & \mzero & \mzero \\
        \mzero & \mzero & \mzero & \mzero & \mzero & \mzero & \mzero & \mzero & \mzero & \mzero & \mzero & \mzero & \mzero & \mzero & \mzero \\
        \mzero & \mzero & \mzero & \mzero & \mzero & \mzero & \mzero & \mzero & \mzero & \mzero & \mzero & \mzero & \mzero & \mzero & \mzero \\
        \mzero & \mzero & \mzero & \mzero & \mzero & \mzero & \mzero & \mzero & \mzero & \mzero & \mzero & \mzero & \mzero & \mzero & \mzero \\
        \mzero & \mzero & \mzero & \mzero & \mzero & \mzero & \mzero & \mzero & \mzero & \mzero & \mzero & \mzero & \mzero & \mzero & \mzero \\
        \mzero & \frac{3}{4} & \mzero & \mzero & -\frac{9}{8} & 1 & \mzero & \mzero & \mzero & \mzero & \mzero & \mzero & \mzero & \mzero & \mzero \\
        \mzero & \mzero & \mzero & \mzero & \mzero & \mzero & \frac{1}{2} & \mzero & \mzero & \mzero & \mzero & \mzero & \mzero & \mzero & \mzero \\
        \mzero & \mzero & \mzero & \mzero & \mzero & \mzero & \mzero & \frac{1}{2} & \mzero & \mzero & \mzero & \mzero & \mzero & \mzero & \mzero \\
        \mzero & \mzero & \mzero & \mzero & \mzero & \mzero & \mzero & \mzero & \mzero & -1 & \mzero & \mzero & \mzero & \mzero & \mzero \\
        \mzero & \mzero & \mzero & \mzero & \mzero & \mzero & \mzero & \mzero & \mzero & 1 & \mzero & \mzero & \mzero & \mzero & \mzero \\
        -\frac{1}{16} & \frac{1}{32} & -\frac{1}{32} & \mzero & \mzero & -\frac{1}{4} & \mzero & \mzero & \mzero & \mzero & 2 & \frac{1}{8} & \mzero & \mzero & \mzero \\
        \frac{1}{2} & \frac{5}{4} & \frac{1}{4} & \mzero & -\frac{9}{4} & 4 & \mzero & \mzero & \mzero & \mzero & -16 & -1 & \mzero & \mzero & \mzero \\
        -\frac{3}{8} & \mzero & -\frac{15}{64} & -\frac{5}{32} & \mzero & \mzero & \mzero & \mzero & 1 & \mzero & -\frac{5}{2} & \mzero & 3 & \frac{1}{2} & \frac{1}{2} \\
        \frac{3}{4} & \mzero & \frac{15}{32} & \frac{5}{16} & \mzero & \mzero & \mzero & \mzero & -2 & 2 & 1\mzero & \mzero & -8 & -1 & -2 \\
        \frac{7}{16} & \frac{5}{32} & \frac{5}{16} & \frac{5}{16} & \mzero & -\frac{5}{4} & \mzero & \mzero & -2 & \mzero & 1\mzero & \frac{5}{8} & -4 & -1 & \mzero
    \end{smallmatrix}\right], \quad A_{-1} = \left[\begin{smallmatrix}
        \mzero & \mzero & \mzero & \mzero & \mzero & \mzero & \mzero & \mzero & \mzero & \mzero & \mzero & \mzero & \mzero & \mzero & \mzero \\
        \mzero & \mzero & \mzero & \mzero & \mzero & \mzero & \mzero & \mzero & \mzero & \mzero & \mzero & \mzero & \mzero & \mzero & \mzero \\
        \mzero & \mzero & \mzero & \mzero & \mzero & \mzero & \mzero & \mzero & \mzero & \mzero & \mzero & \mzero & \mzero & \mzero & \mzero \\
        \mzero & \mzero & \mzero & \mzero & \mzero & \mzero & \mzero & \mzero & \mzero & \mzero & \mzero & \mzero & \mzero & \mzero & \mzero \\
        \mzero & \mzero & \mzero & \mzero & \mzero & \mzero & \mzero & \mzero & \mzero & \mzero & \mzero & \mzero & \mzero & \mzero & \mzero \\
        \mzero & \frac{3}{4} & \mzero & \mzero & \frac{9}{8} & 1 & \mzero & \mzero & \mzero & \mzero & \mzero & \mzero & \mzero & \mzero & \mzero \\
        \mzero & \mzero & \mzero & \mzero & \mzero & \mzero & \frac{1}{2} & \mzero & \mzero & \mzero & \mzero & \mzero & \mzero & \mzero & \mzero \\
        \mzero & \mzero & \mzero & \mzero & \mzero & \mzero & \mzero & \frac{1}{2} & \mzero & \mzero & \mzero & \mzero & \mzero & \mzero & \mzero \\
        \mzero & \mzero & \mzero & \mzero & \mzero & \mzero & \mzero & \mzero & 4 & 1 & \mzero & \mzero & \mzero & \mzero & \mzero \\
        \mzero & \mzero & \mzero & \mzero & \mzero & \mzero & \mzero & \mzero & -12 & -3 & \mzero & \mzero & \mzero & \mzero & \mzero \\
        \frac{1}{16} & -\frac{1}{32} & \frac{1}{32} & \mzero & \mzero & \frac{1}{4} & \mzero & \mzero & \mzero & \mzero & 2 & -\frac{1}{8} & \mzero & \mzero & \mzero \\
        \frac{1}{2} & \frac{5}{4} & \frac{1}{4} & \mzero & \frac{9}{4} & 4 & \mzero & \mzero & \mzero & \mzero & 16 & -1 & \mzero & \mzero & \mzero \\
        -\frac{3}{8} & \mzero & -\frac{15}{64} & -\frac{5}{32} & \mzero & \mzero & \mzero & \mzero & -1 & \mzero & \frac{5}{2} & \mzero & 3 & \frac{1}{2} & \frac{1}{2} \\
        \frac{3}{4} & \mzero & \frac{15}{32} & \frac{5}{16} & \mzero & \mzero & \mzero & \mzero & 1\mzero & 2 & -1\mzero & \mzero & -8 & -1 & -2 \\
        \frac{7}{16} & \frac{5}{32} & \frac{5}{16} & \frac{5}{16} & \mzero & -\frac{5}{4} & \mzero & \mzero & 2 & \mzero & -1\mzero & \frac{5}{8} & -4 & -1 & \mzero \\
    \end{smallmatrix}\right] \nonumber \, ,\\
    A_{0} = \left[\begin{smallmatrix}
        \mzero & \mzero & \mzero & \mzero & \mzero & \mzero & \mzero & \mzero & \mzero & \mzero & \mzero & \mzero & \mzero & \mzero & \mzero \\
         \mzero & \mzero & \mzero & \mzero & \mzero & \mzero & \mzero & \mzero & \mzero & \mzero & \mzero & \mzero & \mzero & \mzero & \mzero \\
         \mzero & \mzero & \mzero & \mzero & \mzero & \mzero & \mzero & \mzero & \mzero & \mzero & \mzero & \mzero & \mzero & \mzero & \mzero \\
         \mzero & \mzero & \mzero & \mzero & \mzero & \mzero & \mzero & \mzero & \mzero & \mzero & \mzero & \mzero & \mzero & \mzero & \mzero \\
        \mzero & \mzero & \mzero & \mzero & -6 & \mzero & \mzero & \mzero & \mzero & \mzero & \mzero & \mzero & \mzero & \mzero & \mzero \\
        \mzero & -\frac{3}{2} & \mzero & \mzero & \mzero & -6 & \mzero & \mzero & \mzero & \mzero & \mzero & \mzero & \mzero & \mzero & \mzero \\
        \mzero & \mzero & \mzero & \mzero & \mzero & \mzero & -3 & \mzero & \mzero & \mzero & \mzero & \mzero & \mzero & \mzero & \mzero \\
        \mzero & \mzero & \mzero & \mzero & \mzero & \mzero & \mzero & -3 & \mzero & \mzero & \mzero & \mzero & \mzero & \mzero & \mzero \\
        \mzero & \mzero & \mzero & \mzero & \mzero & \mzero & \mzero & \mzero & -6 & \mzero & \mzero & \mzero & \mzero & \mzero & \mzero \\
        \frac{5}{4} & \frac{5}{4} & \frac{5}{4} & \frac{5}{4} & \mzero & 5 & \mzero & \mzero & 16 & 2 & \mzero & \mzero & \mzero & \mzero & \mzero \\
        \mzero & \mzero & \mzero & \mzero & \mzero & \mzero & \mzero & \mzero & \mzero & \mzero & -6 & \mzero & \mzero & \mzero & \mzero \\
        1 & -\frac{1}{2} & \frac{1}{2} & \mzero & \mzero & \mzero & \mzero & \mzero & \mzero & \mzero & \mzero & 2 & \mzero & \mzero & \mzero \\
        \mzero & \mzero & \mzero & \mzero & \mzero & \mzero & \mzero & \mzero & \mzero & \mzero & \mzero & \mzero & -6 & \mzero & \mzero \\
        \frac{3}{2} & \frac{5}{4} & \frac{15}{16} & \frac{5}{8} & \mzero & 5 & \mzero & \mzero & \mzero & \mzero & \mzero & \mzero & 16 & 2 & \mzero \\
        \frac{7}{8} & \frac{15}{16} & \frac{5}{8} & \frac{5}{8} & \mzero & \frac{5}{2} & \mzero & \mzero & \mzero & \mzero & \mzero & -\frac{5}{4} & 8 & 2 & \mzero \\
    \end{smallmatrix}\right], \quad A_{w_{0}}(x) = \left[\begin{smallmatrix}
        \mzero & \mzero & \mzero & \mzero & \mzero & \mzero & \mzero & \mzero & \mzero & \mzero & \mzero & \mzero & \mzero & \mzero & \mzero \\
        \mzero & \mzero & \mzero & \mzero & \mzero & \mzero & \mzero & \mzero & \mzero & \mzero & \mzero & \mzero & \mzero & \mzero & \mzero \\
        \mzero & \mzero & \mzero & \mzero & \mzero & \mzero & \mzero & \mzero & \mzero & \mzero & \mzero & \mzero & \mzero & \mzero & \mzero \\
        \mzero & \mzero & \mzero & \mzero & \mzero & \mzero & \mzero & \mzero & \mzero & \mzero & \mzero & \mzero & \mzero & \mzero & \mzero \\
        \mzero & \mzero & \mzero & \mzero & \mzero & \mzero & \mzero & \mzero & \mzero & \mzero & \mzero & \mzero & \mzero & \mzero & \mzero \\
        \mzero & \mzero & \mzero & \mzero & \mzero & \mzero & \mzero & \mzero & \mzero & \mzero & \mzero & \mzero & \mzero & \mzero & \mzero \\
        \mzero & \mzero & \mzero & \mzero & \mzero & \mzero & \mzero & \frac{1}{(x-1) x (x+1) w_0(x)^2} & \mzero & \mzero & \mzero & \mzero & \mzero & \mzero & \mzero \\
        \mzero & \mzero & \mzero & \mzero & -\frac{45}{4} w_0(x) & \mzero & -\frac{\left(8 x^2-9\right) w_0(x)^2}{(x-1) x (x+1)} & \mzero & \mzero & \mzero & \mzero & \mzero & \mzero & \mzero & \mzero \\
        \mzero & \mzero & \mzero & \mzero & \mzero & \mzero & \mzero & \mzero & \mzero & \mzero & \mzero & \mzero & \mzero & \mzero & \mzero \\
        \mzero & \mzero & \mzero & \mzero & \mzero & \mzero & \frac{4 w_0(x)}{x} & \mzero & \mzero & \mzero & \mzero & \mzero & \mzero & \mzero & \mzero \\
        \mzero & \mzero & \mzero & \mzero & \mzero & \mzero & \mzero & \mzero & \mzero & \mzero & \mzero & \mzero & \mzero & \mzero & \mzero \\
        \mzero & \mzero & \mzero & \mzero & \mzero & \mzero & \mzero & \mzero & \mzero & \mzero & \mzero & \mzero & \mzero & \mzero & \mzero \\
        \mzero & \mzero & \mzero & \mzero & \mzero & \mzero & \mzero & \mzero & \mzero & \mzero & \mzero & \mzero & \mzero & \mzero & \mzero \\
        \mzero & \mzero & \mzero & \mzero & \mzero & \mzero & \mzero & \mzero & \mzero & \mzero & \mzero & \mzero & \mzero & \mzero & \mzero \\
        \mzero & \mzero & \mzero & \mzero & \mzero & \mzero & \mzero & \mzero & \mzero & \mzero & \mzero & \mzero & \mzero & \mzero & \mzero \\
    \end{smallmatrix}\right] \, .
\end{align}
\section{Boundary vector}
\label{app:boundary}
In this appendix, we provide the explicit results for the boundary master integrals studied in Sec.~\ref{ssec:integral_evaluation}.
\paragraph{Hard region.}
\begin{figure}[H]
\centering
\includegraphics[width=0.15\textwidth]{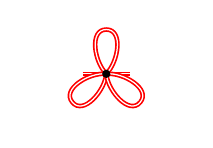}
\includegraphics[width=0.15\textwidth]{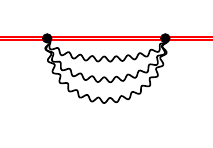}
\includegraphics[width=0.15\textwidth]{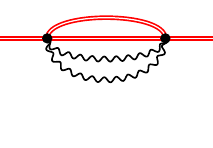}
\includegraphics[width=0.15\textwidth]{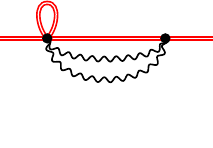}
\includegraphics[width=0.15\textwidth]{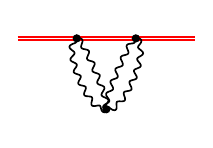}
\includegraphics[width=0.15\textwidth]{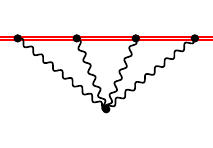}
   \caption{6 master integrals of the hard boundary region.}
  \label{fig:hard masters}
\end{figure}
The first five boundary master integrals in the hard region, Eq.~\eqref{eq:mis_hard}, read as:
\begin{align}
    &F^{\rm h \, (\eta_1 \eta_2 \eta_3)}_{111000000}=\mu^{6 \epsilon} \, e^{3 \epsilon \gamma_E} \, \omega^{3-6 \epsilon} \, \left[\frac{e^{i\pi\left(\eta_1+\eta_2+\eta_3\right)\left(\epsilon-\frac{1}{2}\right)}}{(4 \pi)^{\frac{9}{2}}} \, \Gamma^3\left(\epsilon-\frac{1}{2}\right)\right] \, ,
    \\
    &
    F^{\rm h \, (\eta_1 \eta_2 \eta_3)}_{100001110}=\mu^{6 \epsilon} \, e^{3 \epsilon \gamma_E} \, \omega^{1-6\epsilon} \, \left[ \frac{e^{i \pi \eta_1\left(3\epsilon-\frac{1}{2}\right)}}{(4 \pi)^{\frac{9}{2}}} \, \frac{\Gamma^3\left(\frac{1}{2}-\epsilon\right)\Gamma(2 \epsilon)\Gamma\left(3\epsilon-\frac{1}{2}\right)\Gamma(2-6\epsilon)}{\Gamma \left(\frac{3}{2}-3\epsilon\right)\Gamma(2-4\epsilon)}\right] \, ,
    \\
    &
    F^{\rm h \, (\eta_1 \eta_2 \eta_3)}_{101000110}=
    \mu^{6 \epsilon} \, \e^{3\epsilon\gamma_E} \, \omega^{1-6\epsilon}\, \frac{\Gamma^2\!\left(\tfrac12-\epsilon\right)\Gamma^2(2\epsilon)\,\Gamma\!\left(\epsilon+\tfrac12\right)\Gamma\!\left(3\epsilon-\tfrac12\right)}{(4\pi)^{\frac{9}{2}} \,\Gamma(4\epsilon)\,\Gamma\!\left(\tfrac32-\epsilon\right)}\left[\e^{i\pi\eta_3\left(3\epsilon-\frac{1}{2}\right)}\left(\frac{1+\eta_1\eta_3}{2}\right)+\frac{\sin \pi \epsilon}{\cos 2 \pi \epsilon}\left(\frac{1-\eta_1\eta_3}{2}\right)\right] \, , 
    \\
    &
    F^{\rm h \, (\eta_1 \eta_2 \eta_3)}_{110001010}=\mu^{6 \epsilon}\, e^{3 \epsilon \gamma_E} \, \omega^{1-6\epsilon} \, \left[\frac{e^{i \pi \left[2\epsilon \,\eta_2-\eta_1\left(\frac{1}{2}-\epsilon\right)\right]}}{(4 \pi)^{\frac{9}{2}}} \,\frac{\Gamma^2\left(\frac{1}{2}-\epsilon\right)\Gamma\left(\frac{1}{2}+\epsilon\right)\Gamma\left(2 \epsilon\right)\Gamma\left(1-4 \epsilon\right)\Gamma\left( \epsilon-\frac{1}{2}\right)}{\Gamma\left(1-2\epsilon\right)\Gamma\left(\frac{3}{2}-3\epsilon\right)}\right] \, ,
    \\
    &
    F^{\rm h \, (\eta_1 \eta_2 \eta_3)}_{010101110}= \mu^{6 \epsilon} \, e^{3 \epsilon \gamma_E} \, \omega^{-1-6\epsilon} \left[ \frac{e^{i\pi \eta_2 \left(\frac{1}{2}+3\epsilon\right)}}{\,(4\pi)^{\frac{9}{2}}\,}\frac{\Gamma^2\left(\frac{1}{2}+\epsilon\right)\Gamma^4\left(\frac{1}{2}-\epsilon\right)\Gamma\left(\frac{1}{2}+3\epsilon\right)\Gamma\left(-6\epsilon\right)}{\Gamma^2\left(1-2\epsilon\right)\Gamma\left(1-4\epsilon\right)} \right] \, .
\end{align}
Such expressions were derived using a combination of methods: integrals 1,2,4, and 5 were computed by recursively applying formulas for massless bubble and tadpole integrals, while integral 3 was obtained via Mellin–Barnes techniques.
Integral 6 was obtained by reconstructing the numerical result obtained up to $O(\epsilon^2)$ using \textsc{AMFlow}, for active propagators and for the various cut configurations, giving:
\begin{align}
F^{\rm h \, (+++)}_{111101110}
& 
=i \mu^{6 \epsilon} \, \omega^{-5-6\epsilon} \,
    \frac{e^{3 i \epsilon \pi}}{(16 \pi)^3} \,\biggl[\frac{1}{3\epsilon^2}-\frac{7+6\log(2)}{3\epsilon}+15+14 \log(2)+\frac{29 \pi ^2}{36}+6 \log^2(2)
\nonumber\\ & 
\qquad\qquad \qquad 
    +\left(31 \zeta (3)-93-\frac{1}{12} \pi ^2 (57+58 \log (2))-6 \log (2) (15+\log (2) (7+2 \log
   (2)))\right)\epsilon+\mathcal{O}(\epsilon^2)\biggr] \, ,
   \\ 
F^{\rm h \, (ccc)}_{111101110}
& 
= -\frac{ \mu^{6 \epsilon} \, \omega^{-5-6\epsilon}}{(16\pi)^3}\biggl[\frac{4}{3\epsilon^2}
-\frac{1}{\epsilon}\left(\frac{44}{3}+8\log 2\right)
+\left(\frac{316}{3}+88\log 2+24\log^2 2-\frac{241\pi^2}{9}\right)
\nonumber \\
& \quad
+\epsilon\left(
-\frac{6092}{9}
-632\log 2
-264\log^2 2
-48\log^3 2
+\frac{\pi^2}{9}\left(2203+1446\log 2\right)
+124\zeta_3
\right) \biggr]  +\mathcal{O}(\epsilon^2) \, ,
\\
F^{\rm h \, (cc+)}_{111101110}
& 
= -\frac{i \mu^{6 \epsilon} \, \omega^{-5-6\epsilon}}{(16\pi)^3}\biggl[
-\frac{2}{3\epsilon^2}
+\frac{2}{3\epsilon}\left(11+i\pi+6\log2\right)
+\frac{1}{18}\left[
241\pi^2
-12\left(79+66\log2+18\log^2 2\right)
-12i\pi\left(5+6\log2\right)
\right]\biggr]
\nonumber \\
& \quad
+\frac{\epsilon}{18}\left[
4\left(1523+1422\log2+594\log^2 2+108\log^3 2-279\zeta_3\right)
-\pi^2\left(2203+1446\log2\right)
\right.
\nonumber \\
& \quad\left.
+i\pi\left(3\pi^2+36\left(9+10\log2+6\log^2 2\right)\right)
\right] \biggr] +\mathcal{O}(\epsilon^2) \, ,
\\
F^{\rm h \, (c+c)}_{111101110}
& = -\frac{i \mu^{6 \epsilon} \, \omega^{-5-6\epsilon}}{(16\pi)^3}\biggl[
-\frac{2}{3\epsilon^2}
+\frac{2}{3\epsilon}\left(11-5i\pi+6\log2\right)
+\frac{1}{18}\left[
241\pi^2
+2i\pi\left(407+180\log2\right)
-12\left(79+66\log2+18\log^2 2\right)
\right]
\nonumber \\
& \quad
+\frac{\epsilon}{18}\left[
4\left(1523+1422\log2+594\log^2 2+108\log^3 2-279\zeta_3\right)
-\pi^2\left(2203+1446\log2\right)
\right.
\nonumber \\
& \quad\left.
+i\pi\left(
513\pi^2
-12\left(319+282\log2+90\log^2 2\right)
\right)
\right] \biggr]  +\mathcal{O}(\epsilon^2)\, , 
\\
F^{\rm h \, (+cc)}_{111101110} & = F^{\rm h \, (cc+)}_{111101110} \,.
\end{align}

\paragraph{Soft region.}
\begin{figure}[H]
\centering
\includegraphics[width=0.15\textwidth]{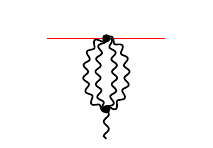}
\includegraphics[width=0.15\textwidth]{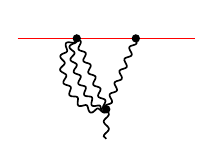}
\includegraphics[width=0.15\textwidth]{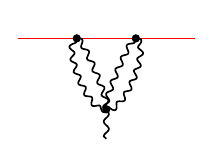}
\includegraphics[width=0.15\textwidth]{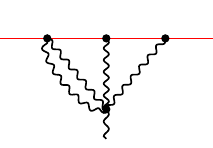}
\includegraphics[width=0.15\textwidth]{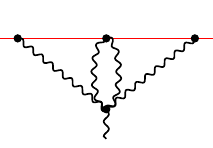}
\includegraphics[width=0.15\textwidth]{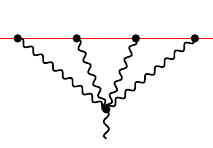}
   \caption{6 master integrals of the soft boundary region.}
  \label{fig:soft masters}
\end{figure}
The boundary master integrals in the soft region, Eq.~\eqref{eq:mis_soft}, read as:
\begin{align}
    F^{\rm s \, (\eta_1 \eta_2 \eta_3)}_{000100001110}&= \mu^{6\epsilon} \, e^{3 \epsilon \gamma_E} \, \omega^{1-6\epsilon}\left[\frac{1}{(4 \pi)^{\frac{9}{2}}}\frac{\Gamma^4 \left(\frac{1}{2}-\epsilon\right) \Gamma \left(3 \epsilon-\frac{1}{2}\right)}{2^{6 \epsilon-1} \, \Gamma (2-4 \epsilon)}\right] \, ,\\
    F^{\rm s \, (\eta_1 \eta_2 \eta_3)}_{001100001110}&=\mu^{6 \epsilon} \, e^{3 \epsilon \gamma_E} \, \omega^{-1-6\epsilon}\left[\frac{e^{\frac{i \pi}{2}\eta_3}}{(4\pi)^{\frac{9}{2}}}\frac{\sqrt{\pi} \, \Gamma^3\left(\frac{1}{2}-\epsilon\right) \, \Gamma\left(-\epsilon\right)\Gamma\left(1-3\epsilon\right)\Gamma\left(3\epsilon\right)}{2^{6 \epsilon} \, \Gamma\left(\frac{3}{2}-3\epsilon\right)\Gamma\left(1-4\epsilon\right)}\right] \, ,\\
    F^{\rm s \, (\eta_1 \eta_2 \eta_3)}_{010100001110}&=\mu^{6 \epsilon} \, e^{3 \epsilon \gamma_E} \, \omega^{-1-6\epsilon}\left[\frac{e^{\frac{i \pi}{2}\eta_2}}{(4\pi)^{\frac{9}{2}}}\frac{\sqrt{\pi} \, \Gamma^4\left(\frac{1}{2}-\epsilon\right) \, \Gamma^2\left(\frac{1}{2}-2\epsilon\right)\Gamma\left(3\epsilon\right)}{2^{6 \epsilon} \, \Gamma^2\left(1-2\epsilon\right)\Gamma\left(1-4\epsilon\right)}\right] \, ,\\
    F^{\rm s \, (\eta_1 \eta_2 \eta_3)}_{011100001110}&=\mu^{6 \epsilon} \, e^{3 \epsilon \gamma_E} \, \omega^{-3-6\epsilon}\Bigg\{ \frac{\eta_2 \, \eta_3}{(4\pi)^{\frac{9}{2}}}\frac{\Gamma^2\!\left(\frac12-\epsilon\right)}{2^{6 \epsilon +1} \, \Gamma(1-2\epsilon)}\Bigg[-\frac{\pi\,\Gamma\!\left(\frac12-2\epsilon\right)\Gamma^2(-\epsilon)\Gamma\!\left(\frac12+3\epsilon\right)}{\Gamma\!\left(\frac12-4\epsilon\right)}- \eta_2 \, \eta_3 \,\Gamma\!\left(\frac32-\epsilon\right)\Gamma\!\left(-\frac12+\epsilon\right) \times \nonumber \\
    & \times \Bigg(\Gamma\!\left(\frac52-3\epsilon\right)\Gamma(-2\epsilon)\Gamma(-\epsilon)\Gamma\!\left(-\frac32+3\epsilon\right)\,{}_3\widetilde{F}_2\!\left(\begin{matrix}-\epsilon,\,-2\epsilon,\,\frac12-3\epsilon\\[2mm]1-\epsilon,\,\frac12-5\epsilon\end{matrix};1\right)  \nonumber \\
    &-\sqrt{\pi}\,\Gamma(1-2\epsilon)\Gamma\!\left(\frac12-\epsilon\right)\Gamma\!\left(\frac12+3\epsilon\right)\,{}_4\widetilde{F}_3\!\left(\begin{matrix}\frac12,\,1,\,\frac12-\epsilon,\,1-2\epsilon\\[2mm]\frac32,\,\frac32+\epsilon,\,1-4\epsilon\end{matrix};1\right)\Bigg)\Bigg]\Bigg\} \, ,
    \\
    F^{\rm s \, (\eta_1 \eta_2 \eta_3)}_{101100001110}&=\mu^{6 \epsilon} \, e^{3 \epsilon \gamma_E} \, \omega^{-3-6\epsilon}\Bigg\{\frac{\eta_1 \, \eta_3}{(4\pi)^{\frac{9}{2}}}\frac{\Gamma^2\!\left(\frac12-\epsilon\right)}{2^{6 \epsilon +1} \, \Gamma(1-2\epsilon)}\Bigg[-\frac{\pi\,\Gamma\!\left(\frac12-2\epsilon\right)\Gamma^2(-\epsilon)\Gamma\!\left(\frac12+3\epsilon\right)}{\Gamma\!\left(\frac12-4\epsilon\right)}+ \eta_1 \, \eta_3 \, \Gamma(1-2\epsilon)\,\Gamma(2\epsilon)\,\Gamma\!\left(\frac12+3\epsilon\right) \times \nonumber \\& \times\Bigg(\Gamma^2\!\left(\frac12-3\epsilon\right)\Gamma\!\left(\frac12-2\epsilon\right)\,{}_3\widetilde{F}_2\!\left(\begin{matrix}\frac12-2\epsilon,\,\frac12-3\epsilon,\,\frac12-3\epsilon\\[2mm]1-6\epsilon,\,\frac32-2\epsilon\end{matrix};1\right)\nonumber \\&-\sqrt{\pi}\,\Gamma^2\!\left(\frac12-\epsilon\right)\,{}_4\widetilde{F}_3\!\left(\begin{matrix}\frac12,\,1,\,\frac12-\epsilon,\,\frac12-\epsilon\\[2mm]\frac32,\,1-4\epsilon,\,1+2\epsilon\end{matrix};1\right)\Bigg)\Bigg]\Bigg\} \, , \\
    F^{\rm s \, (\eta_1 \eta_2 \eta_3)}_{111100001110}&=\mu^{6 \epsilon} \, e^{3 \epsilon \gamma_E} \, \omega^{-5-6\epsilon} \left(\frac{\eta_1+\eta_3}{3}-\eta_1 \eta_2 \eta_3\right)\left[\frac{1}{(4\pi)^{\frac{9}{2}}}\frac{i \, \pi ^{3/2} \Gamma^4 (-\epsilon) \, \Gamma (3 \epsilon+1)}{ 2^{6 \epsilon+2} \, \Gamma (-4 \epsilon)}\right] \, ,
\end{align}
where we used
\begin{equation}
    {}_p\widetilde{F}_q\!\left(\vec a;\vec b;z\right)=\frac{{}_pF_q\!\left(\vec a;\vec b;z\right)}{\Gamma(b_1)\cdots\Gamma(b_q)} \, .
\end{equation}
The first three integrals are derived analytically using standard massless bubble and tadpole integrals together with the result of~\cite{Dlapa:2023hsl,Smirnov:2012gma} 
\begin{equation}
    \int _{\boldsymbol{\ell}} \frac{1}{[\pm \, \boldsymbol{\ell} \cdot \boldsymbol{n}-i0^+]^{\alpha_1} \, [\boldsymbol{\ell}^2]^{\beta_1}[(\boldsymbol{\ell}-\boldsymbol{\hat{q}})^2]^{\beta_2}}=\frac{e^{\epsilon\gamma_E}}{(4\pi)^{\frac{3}{2}}}\frac{2^{\alpha_1-1}i^{\alpha_1}\Gamma\!\left(\frac{\alpha_1}{2}\right)\Gamma\!\left(\frac{d-\alpha_1}{2}-\beta_1\right)\Gamma\!\left(\frac{d-\alpha_1}{2}-\beta_2\right)\Gamma\!\left(\frac{\alpha_1-d}{2}+\beta_1+\beta_2\right)}{\Gamma(\alpha_1)\,\Gamma(\beta_1)\,\Gamma(\beta_2)\,\Gamma(d-\alpha_1-\beta_1-\beta_2)} \, ,
    \label{eq:master_boundary}
\end{equation}
where $d=D-1$, $\beta_1>0$, $\beta_2>0$, $\boldsymbol{n}\cdot\boldsymbol{\hat{q}}=0$, and $\boldsymbol{\hat{q}}^2=1$. These derivations were generalized to arbitrary prescriptions. The fourth and fifth integrals require a dedicated derivation based on a combination of Schwinger and Feynman parameterizations. The full derivation of these integrals can be found in Refs.~\cite{Dlapa:2023hsl,Jinno:2022sbr}, together with the computation of the sixth top-sector integral. These results have been verified numerically using $\textsc{AMFlow}$.
\end{document}